\documentclass[twocolumn,nofootinbib]{aastex63}
\usepackage{graphicx}	
\usepackage{amsmath}
\usepackage{amssymb}
\usepackage{subfigure}
\usepackage{color}
\usepackage{txfonts}
\usepackage{braket}
\usepackage{multirow}
\usepackage{hyperref,graphicx}
\usepackage{times}
\usepackage{comment}
\usepackage{makecell}
\usepackage{CJK}

\newcommand\simlt{\lower.5ex\hbox{$\; \buildrel < \over \sim \;$}}
\newcommand\simgt{\lower.5ex\hbox{$\; \buildrel > \over \sim \;$}}

\begin{document}
\begin{CJK*}{UTF8}{gbsn}

\title{The Imprint of Large Scale Structure on the Ultra-High-Energy Cosmic Ray Sky}

\author{Chen Ding (丁忱)}
\altaffiliation{Email: cd2209@nyu.edu}
\affiliation{Center for Cosmology and Particle Physics, New York University, New York, NY 10003, USA}

\author{No\'emie Globus}
\altaffiliation{Email: nglobus-visitor@flatironinstitute.org and noemie.globus@eli-beams.eu Current address: ELI Beamlines, Institute of Physics, Czech Academy of Sciences, 25241 Dolni Brezany, Czech Republic.}
\affiliation{Center for Cosmology and Particle Physics, New York University, New York, NY 10003, USA}
\affiliation{Center for Computational Astrophysics, Flatiron Institute, Simons Foundation, New York, NY 10010, USA}

\author{Glennys R. Farrar}
\altaffiliation{Email: gf25@nyu.edu}
\affiliation{Center for Cosmology and Particle Physics, New York University, New York, NY 10003, USA}


\begin{abstract}
{Ultra-high-energy cosmic rays (UHECRs) are atomic nuclei from space with vastly higher energies than any other particles ever observed. Their origin and chemical composition remain a mystery.  As we show here, the large- and intermediate-angular-scale anisotropies observed by the Pierre Auger Observatory are a powerful tool for understanding the origin of UHECRs. Without specifying any particular production mechanism, but only postulating that the source distribution follows the matter distribution of the local Universe, a good accounting of the magnitude, direction and energy dependence of the dipole anisotropy at energies above $8 \times 10^{18}$ eV is obtained, after taking into account the impact of energy losses during propagation (the ``GZK horizon''), diffusion in extragalactic magnetic field and deflections in the Galactic magnetic field (GMF). This is a major step toward the long-standing hope of using UHECR anisotropies to constrain UHECR composition and magnetic fields. The observed dipole anisotropy is incompatible with a pure proton composition in this scenario. With a more accurate treatment of energy losses, it should be possible to further constrain the cosmic-ray composition and properties of the extragalactic magnetic field, self-consistently improve the GMF model, and potentially expose individual UHECR sources. }  

\end{abstract}
\keywords{Cosmic ray astronomy (324); Cosmic ray sources (328); Cosmic rays (329); Extragalactic magnetic fields (507); Milky Way magnetic fields (1057); Ultra-high-energy cosmic radiation (1733); Cosmic anisotropy (316); Large-scale structure of the universe (902)}
\section{Introduction}

Cosmic rays are atomic nuclei that travel through space at almost the speed of light. The existence of cosmic rays of energy beyond $10^{20}$ eV (100 EeV) has been known for nearly 60 years \citep{Linsley1963}, but their origin and composition are still an enigma. The fact that cosmic rays are deflected on their way to the Earth by the extragalactic and Galactic magnetic fields (EGMF and GMF) makes it difficult to pinpoint their sources. Two observatories, the Pierre Auger Observatory \citep{abraham2004properties} and the Telescope Array \citep[TA;][]{Kawai:2008zza}, have made great efforts during the last decade in measuring the energy spectrum, composition and arrival directions of ultra-high-energy cosmic rays (UHECRs).  

A breakthrough in UHECR astrophysics came in 2017,  when the Pierre Auger Observatory reported the first anisotropy to pass the 5$\sigma$ discovery threshold:  a dipole with an amplitude of $6.5^{+1.3}_{-0.9}\%$, with a direction that points far away from the Galactic center \citep{Auger_Science_2017,Auger_all_bins_2020} and an intensity that increases with energy \citep{Auger_4bin_2018,Auger_all_bins_2020}. Auger and TA have also reported observations of anisotropies on intermediate angular scales, with a hotspot above 37 EeV reported by Auger at a 3.9$\sigma$ significance level and a hotspot above 57\footnote{Note there is evidence of an offset in the Auger and TA energy scales, such that 57 EeV for TA corresponds to $\sim$43 EeV for Auger \citep{AugerTA_EPJ}.} EeV reported by TA at 2.9$\sigma$ \citep{ICRC2019_TA_hotspot}. Such observations offer a great opportunity for investigating the origin of UHECRs.

Before presenting our hypothesis, we remind the reader how the perspective on UHECR origins has evolved:  in early days it was assumed that UHECRs were protons, and that magnetic deflections would not be more than a few degrees, so unless individual sources were so weak as to typically contribute zero to one event to the dataset, sources could be identified by clusters of events \citep{Hayashida:1996bc}.  This picture was dislodged due to the absence of significant small-scale clustering as statistics increased, and direct evidence from Auger that protons account for a small fraction of UHECRs $\geq 8$ EeV \citep[see][for a contemporaneous review]{Allard:2011aa}.  Lack of small-scale clustering could be due to a high source density or larger magnetic deflections associated with the higher $Z$ of heavier nuclei; thus, a low source density may still be allowed. 

We adopt here the simplest possible starting hypothesis:  the UHECR source distribution follows the large scale structure (LSS) of the universe.   This must be valid, generally with some bias factor, unless UHECR sources are so rare and powerful that stochastic effects outweigh the mean distribution. The idea of using UHECRs to probe the LSS of the universe was proposed by \citet{waxman1997signature} (see also \citet{Ahlers:2017wpb} for a more recent account). Thanks to work by cosmologists \citep{courtois2013cosmography, Hoffman2018}, the LSS is well-enough known up to a few hundred megaparsecs that we can take it as an input in our model.

The UHECR arrival direction pattern is influenced not only by the source distribution but also by energy losses and photo-disintegration reactions during propagation \citep{GZK-G, GZK-ZK}, and deflections in the EGMFs and GMFs.  
The main uncertainties in predicting the UHECR anisotropy for a given source hypothesis come from our limited knowledge of the EGMFs and GMFs, and our lack of understanding of hadronic physics that leads to uncertainty on the cosmic ray composition. We allow for these uncertainties via parameters in our modeling.

We continue the line of work initiated in \citet{GP17,GPH18}, with three key advances.  First, we account for deflections in the GMF which significantly changes both the magnitude and direction of the dipole anisotropy (bringing both into much better agreement with observations)\footnote{\citet{GPH18} took a preliminary look at the impact of deflections, by tracking $2 \times 10^5$ particles for a single rigidity and coherence length.}.  Second, we systematically treat composition uncertainties, and third, we explore the limitations of the ``sharp cutoff" treatment of \citet{GP17,GPH18} and take first steps toward a full, accurate analysis.   

\begin{figure*}
\includegraphics[width=1\linewidth, trim=0cm 0cm 0cm 0cm, clip=true]{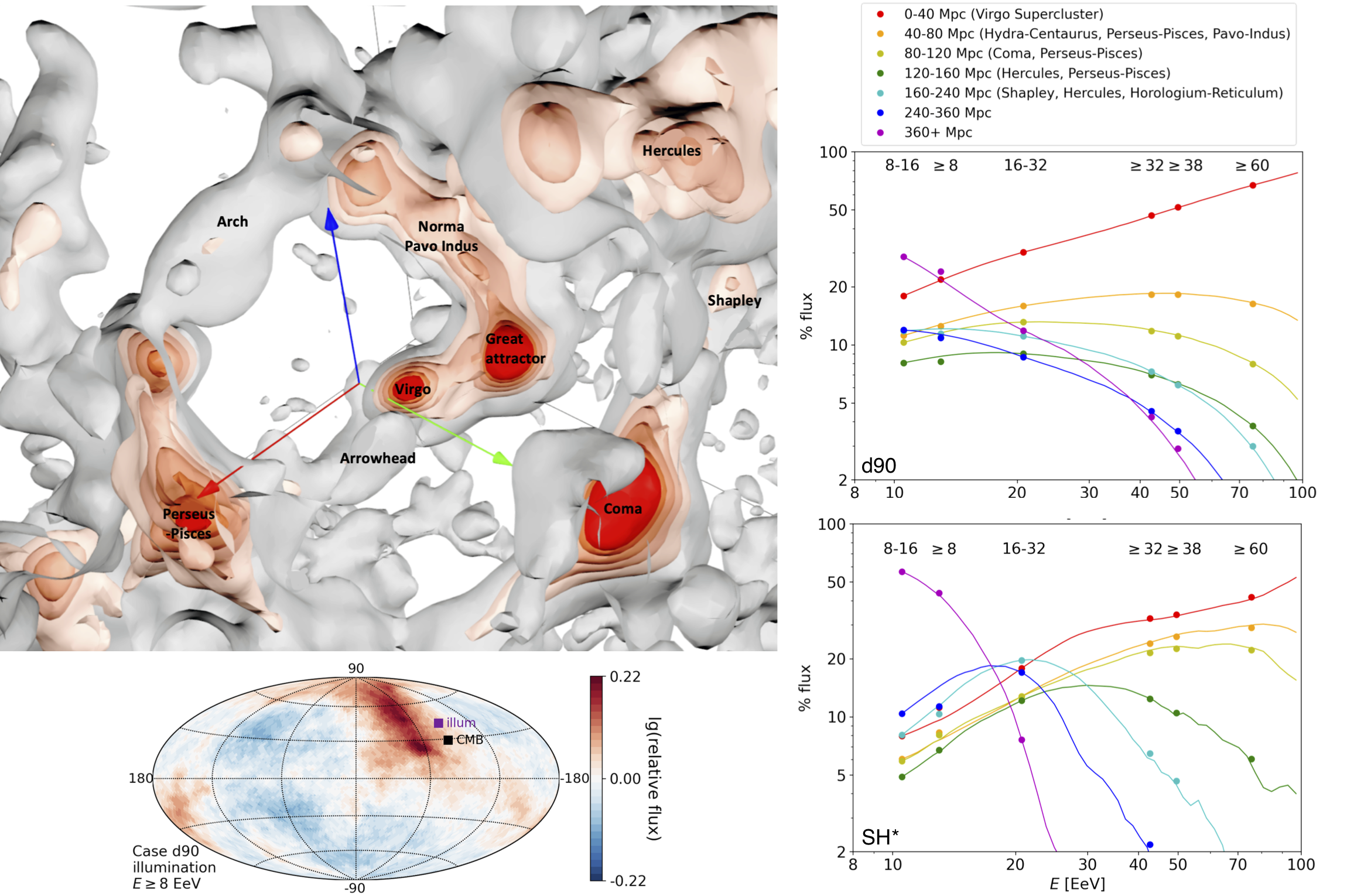}
\caption{Top left: The density field of the local universe derived from {\it CosmicFlow-2} \citep{Hoffman2018} in Supergalactic coordinates;  a 3D interactive view is available at \href{https://sketchfab.com/3d-models/quasi-linear-construction-of-the-density-field-91448f58ed5b4a30b5dc270a34fb4352}{[https://sketchfab.com/3d-models/quasi-linear-construction-of-the-density-field-91448f58ed5b4a30b5dc270a34fb4352]} Bottom left: The intensity map of the flux illuminating the Galaxy $\geq 8$ EeV, for sources following the {\it CosmicFlow-2} density field using the Eq.~\eqref{eq:d90} ``exponential attenuation" treatment; the pattern is virtually identical for the ``sharp-cutoff'' treatment, but with maximum relative flux =1.47 instead of 1.67 as in ``exponential attenuation".  The direction of the dipole component is not far from the CMB dipole.
In the right panels: The colored lines are the percentage contribution to the observed UHECR flux coming from the indicated distance bins, as a function of energy, for the parameters of the best-fitting d90 (top) and SH* (bottom) models detailed in Table~\ref{tab:likelihood}.  The dots represent the average over the energy bin indicated at the top.  The actual calculation uses 1 Mpc bins in distance and 0.02 bins in log$_{10}(E)$.}
\label{fig:figure_1} 
\end{figure*}

In this {\it Letter}, we show that the observed dipole anisotropy and its evolution with energy $\geq 8$ EeV 
are nicely explained as a signature of the local large scale distribution of matter, deflected by GMFs.  This good accounting is obtained if UHECRs have a mixed composition -- consistent with Auger measurements -- but cannot be explained if UHECRs $\geq 8$ EeV are predominantly protons.  These are robust conclusions as they are based on an improved treatment of energy losses during propagation, superceding the sharp cutoff treatment of \citet{GP17,GPH18} which assumed no energy loss out to a ``horizon distance'', then no transmission beyond.  
Our studies suggest that with still more accurate treatments in the future, it should be possible to use anisotropies on large and intermediate scales to simultaneously constrain the composition and potentially improve GMF modeling, as will be elaborated below.  This delineates a future research program with ample rewards. 



\section{Elements of the Analysis}
\subsection{Source distribution model}
We assume the spatial distribution of UHECR sources follows the matter density distribution, and that all sources are ``standard candles"  in terms of luminosity, energy spectrum and composition. We refer to this as the ``LSS model". As in \citet{GPH18}, we use the LSS matter density field \citep{Hoffman2018} derived from the {\it CosmicFlows-2} catalog of peculiar velocities \citep{Tully2014}. 
The 3D distribution of LSS in our local universe is shown in  Fig.~\ref{fig:figure_1}.  With a distance of $\sim17$ Mpc, the Virgo cluster produces a strong excess in the illumination map (IM, the map of flux illuminating the surface of the Milky Way) in a direction close to the Galactic North Pole. The Virgo cluster is part of the Local Supercluster, which contributes most of the excesses within $\sim$40 Mpc \citep{tully2013cosmicflows, pomarede2020cosmicflows}. Beyond the Local Supercluster, the Northern sky is dominated by the Hydra-Centaurus, Coma, Hercules, and Shapley superclusters, while the Southern sky is dominated by the Perseus-Pisces and Horologium-Reticulum superclusters. The further the distance the more isotropic the universe appears, because a given solid angle corresponds to a larger volume of space to average over.  

\subsection{Extragalactic energy loss and magnetic diffusion}\label{subsec:source_horizon}
To take into account horizon effects, we consider two approaches.\vspace{0.1in} \\
1) In the ``sharp cutoff'' approach of \citet{GPH18}, a cosmic-ray is taken to originate with uniform probability from any distance up to a fixed {\it horizon}, $H(A,E)$, where $A$ is the mass number. The sources beyond horizon do not contribute to the flux received at Earth. With no EGMF, the horizon is taken to be the energy attenuation length, $ \chi_{\rm loss}$, also denoted $d_{\rm GZK}$ \citep{GPH18}.  
Diffusion in the EGMF extends the path-length beyond the linear distance to the source and in the presence of an EGMF the horizon is taken to be $\sqrt{ d_{\rm diff} d_{\rm GZK}}$, where $d_{\rm diff}$ is the magnetic diffusion length. For specifics see  \citet{GPH18} and Appendix~\ref{Appendix:illumination}. In the following we refer to this as the ``sharp cutoff'' treatment.  
\vspace{0.1in} \\
2) In our more realistic exponential attenuation treatment, the contribution of a shell at distance $z$ is weighted in its contribution to the observed spectrum in proportion to the factor
\begin{equation}
\label{eq:d90}
\exp\left[- \ln(10) \, p(z,d_{\rm diff}) / d_{90}(A,E)\right],~ 
\end{equation}
where  $p(z,d_{\rm diff})$ is the mean path-length accumulated during traversal of the linear distance $z$ and $d_{90}(A,E)$ is the distance within which 90\% of the parent nuclei of UHECRs in composition bin $A$ and energy above $E$ originated. Since our explorations of EGMF diffusion using the sharp cutoff treatment favor a low or negligible EGMF impact, as discussed below, we restrict ourselves here to taking $p(z,d_{\rm diff})\rightarrow z$. In the following we refer to this as the ``exponential treatment''. It should be noted that this treatment is not an exact description because the energy loss rate evolves during propagation in a non-trivial way as the composition and energy change; a differential description of the evolution in $\{A,\,E\}$ space would be more accurate.  Moreover even if the mean treatment of Eq.~\eqref{eq:d90} is an adequate approximation, the spectrum and composition at the source location should be treated as unknown and varied as part of the fitting procedure -- the spectral and composition sensitivity being reflected in the fact that $d_{90}(A,E)$ depends on these. \citet{Auger_combined_fit} provided the best current combined fit to spectrum and composition at the source, but did not provide corresponding tables of $d_{90}(A,E)$. We use $d_{90}(A,E)$ from \citet{GAP08} which is the most suitable compilation we could find in the literature;  that is based on a spectral index of -2.4, similar to one of the minima found in \citet{Auger_combined_fit}. Given the sensitivity of the \citet{Auger_combined_fit} best fit spectral parameters to the hadronic interaction model (HIM) used for analyzing the composition, the source spectral index must currently be considered quite uncertain. Happily, as discussed in Appendix~\ref{Appendix:horizon}, the analysis is relatively insensitive to this uncertainty.

For either treatment of propagation, we adjust the cumulative contribution of all the distance shells, for each energy and composition bin, to match the observed spectrum at Earth \citep{Auger_PRL_2020,Auger_PRD_2020}.


\begin{figure*}
\centering
    \centering
         \centering
         \includegraphics[width=0.33\linewidth,trim=2cm 39cm 2cm 8cm,clip=true]{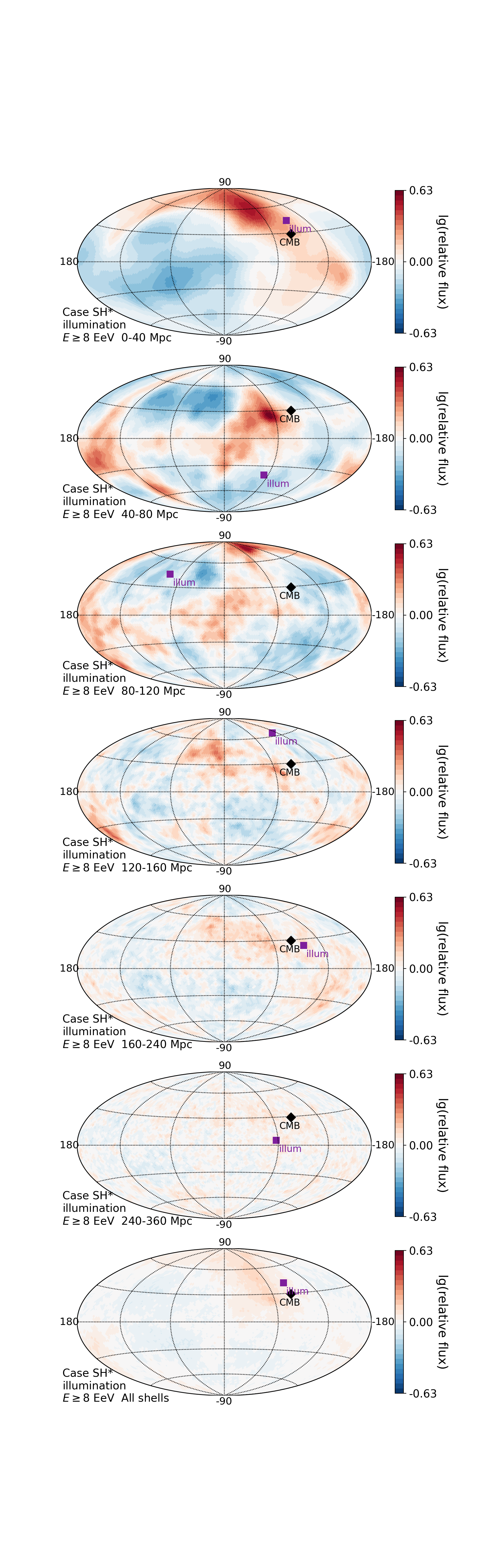}
         \includegraphics[width=0.33\linewidth,trim=2cm 39cm 2cm 8cm,clip=true]{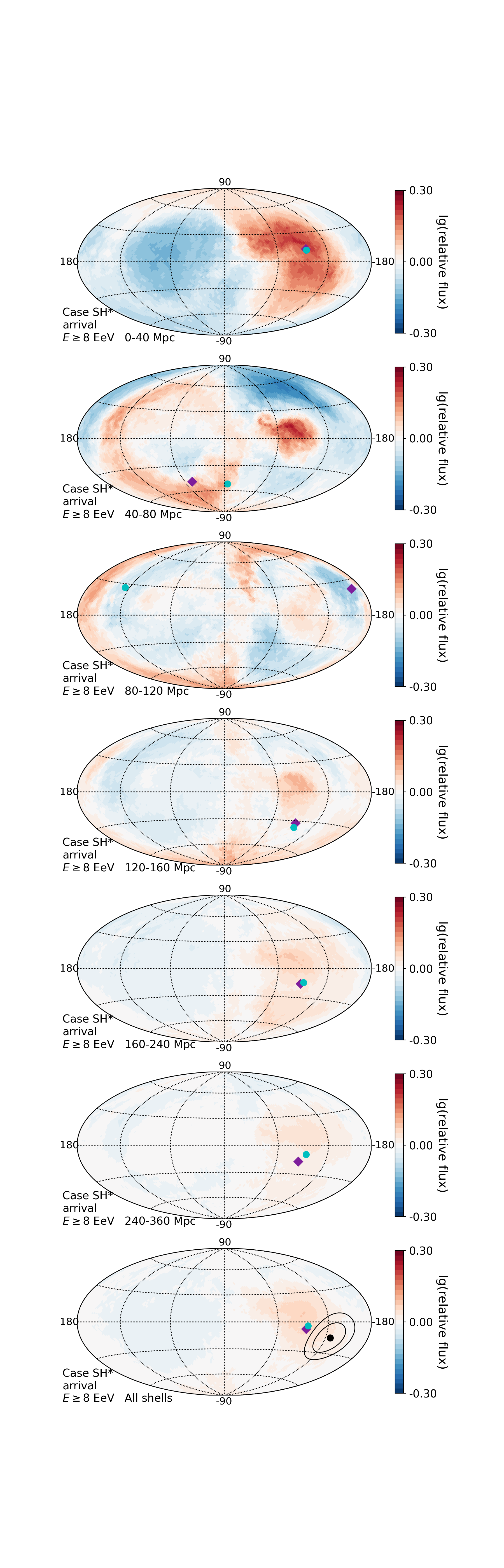}
         \includegraphics[width=0.33\linewidth,trim=2cm 39cm 2cm 8cm,clip=true]{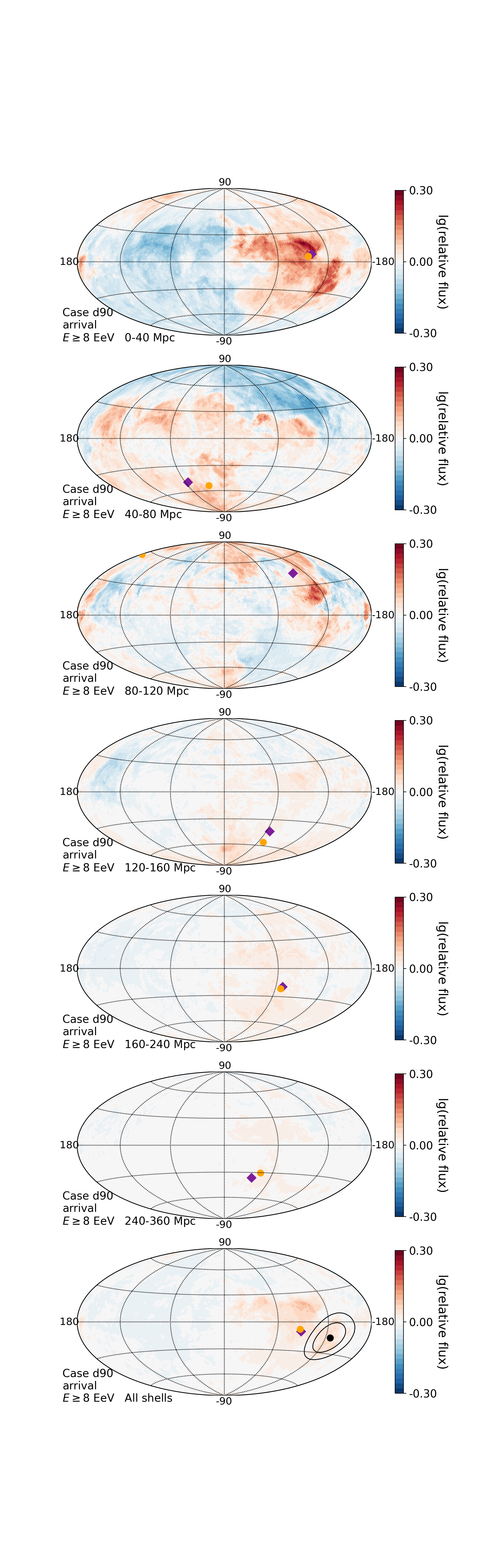}

\caption{Surface density maps (left column), and arrival maps after propagation in the JF12 GMF model for the SH* (middle) and d90 (right) attenuation models, in LSS shells covering distances (top to bottom): 0-40, 40-80, and 80-120 Mpc. The maps here are for $E\geq 8$ EeV;  plots including $E\geq 32$ EeV and a more complete set of distances and models are given in the Appendix Figs.~\ref{fig:allshells_SHasterisk}--~\ref{fig:allshells_PP}.
}
\label{fig:3shells} 
\end{figure*}

We explore the possible spreading of the source images and reduction in horizon due to diffusion in the EGMF, using the sharp cutoff treatment.  We adopt the simplest hypothesis that the universe is filled with homogeneous and isotropic turbulent magnetic fields.  While the turbulence level of the EGMF is still unknown, upper limits obtained by various measurements or arguments exist \citep{durrer2013cosmological}.  We adopt a Kolmogorov spectrum and -- to fully cover the possible parameter space -- we consider rms random field strength $0.08 \leq B_{\rm EG} \leq 10$ nG and coherence length $0.08 \leq \lambda_{\rm EG} \leq 0.5$ Mpc. The diffusion coefficient, $D_{\mathrm EG}$, and indeed all magnetic deflections, depends on rigidity, $E/Z$;  in the relevant rigidity domain, $D_{\mathrm EG}$ is proportional to $\left({E}/{Z B_{\mathrm{EG}} \lambda_{\mathrm{EG}}^{0.5}}\right)^{2}$ \citep{GAP08}. 
The intensity profile of a single source depends on the diffusion coefficient and on the distance to the source; it is calculated by a method following the diffusion of light in scattering media, that allows to take into account the transition between quasi-linear and diffusive regimes, as detailed in Appendix A.  

For a given assumed EGMF, composition and energy, and adopting either the sharp cutoff or the exponential attenuation, we calculate the weight of a 1-Mpc-thick shell of matter at distance~$z$ in the total observed CR flux at the given $(A,E)$.  The final illumination map for that $(A,E)$ and attenuation model is then the weighted sum of the surface mass density in each shell. To enable the reader to visualize how different distances contribute, the left column of Fig.~\ref{fig:3shells} shows the surface density contrast of the three nearest 40-Mpc-thick layers of the nearby universe.  After weighting by the attenuation factor, one gets the contribution of the shell to the illumination map.  The middle and right columns should be ignored for now; they show contributions to the arrival direction maps and will be discussed later. 

\subsection{Galactic propagation}
Once extragalactic cosmic rays enter the Galaxy, they are deflected by the GMF which has a complex geometry and includes both ordered (coherent) and turbulent (random) magnetic fields. The \citet{JF12} model (JF12) is the leading GMF model available at this time. It is constrained by some 40,000 Faraday rotation measures (RMs) of extragalactic sources, and by polarized and unpolarized synchrotron emission as inferred by the Wilkinson Microwave Anisotropy Probe (WMAP).  The JF12 model allows for a coherent and/or turbulent poloidal (X-shaped) halo field inspired by the field geometry seen in external galaxies.   \citep[See][ and Appendix~\ref{Appendix:JF12} for discussion of other GMF models, the limitations of JF12 and references.]{fCRAS14}  The unexpected finding of JF12 was that the X-field is actually a directed poloidal field, extending from South to North. In addition to the X-field, JF12 incorporated spiral arms and a toroidal halo and allowed for the presence of anisotropic random magnetic fields.   

To model the propagation of UHECRs through the Galaxy, we use high-resolution particle tracking (1.8 billion trajectories) from \citet{FS} in the JF12 coherent+random GMF model.  This procedure captures deflection, diffusion and magnification and demagnification effects \citep{FS}.   The coherence length of the GMF is not constrained by the JF12 or other GMF modeling, but it is commonly thought to be 30-100 pc in the bulk of the volume \citep{fCRAS14}.  With that in mind,    \citet{FS} provided trajectories for $\lambda_{\rm coh} = 30$ and 100 pc. Given these high resolution simulated trajectories, the maps of flux at Earth can be calculated from the illumination maps (see Appendix~\ref{Appendix:arrival}).  In order to have more choices of effective GMF coherence lengths $\lambda_{\mathrm{G}}$, we mix the trajectories of 30 parsec and 100 parsec with different weights to interpolate between them.  

\begin{figure*}
\centering
\includegraphics[width=0.26\linewidth, trim=0.2cm 0.2cm 0.7cm 0.6cm, clip=true]{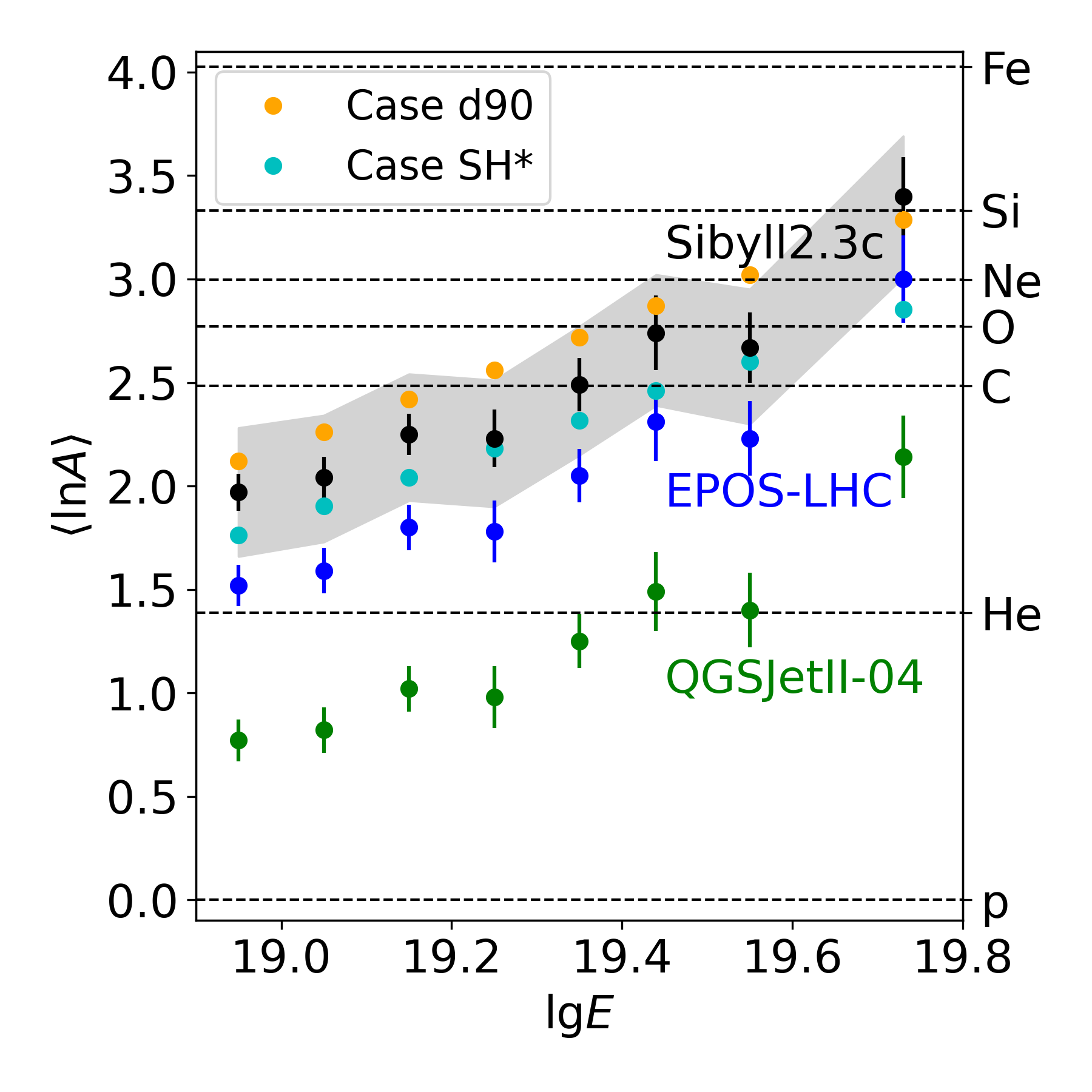}
\includegraphics[width=0.26\linewidth, trim=0.2cm 0.2cm 0.5cm 0.6cm, clip=true]{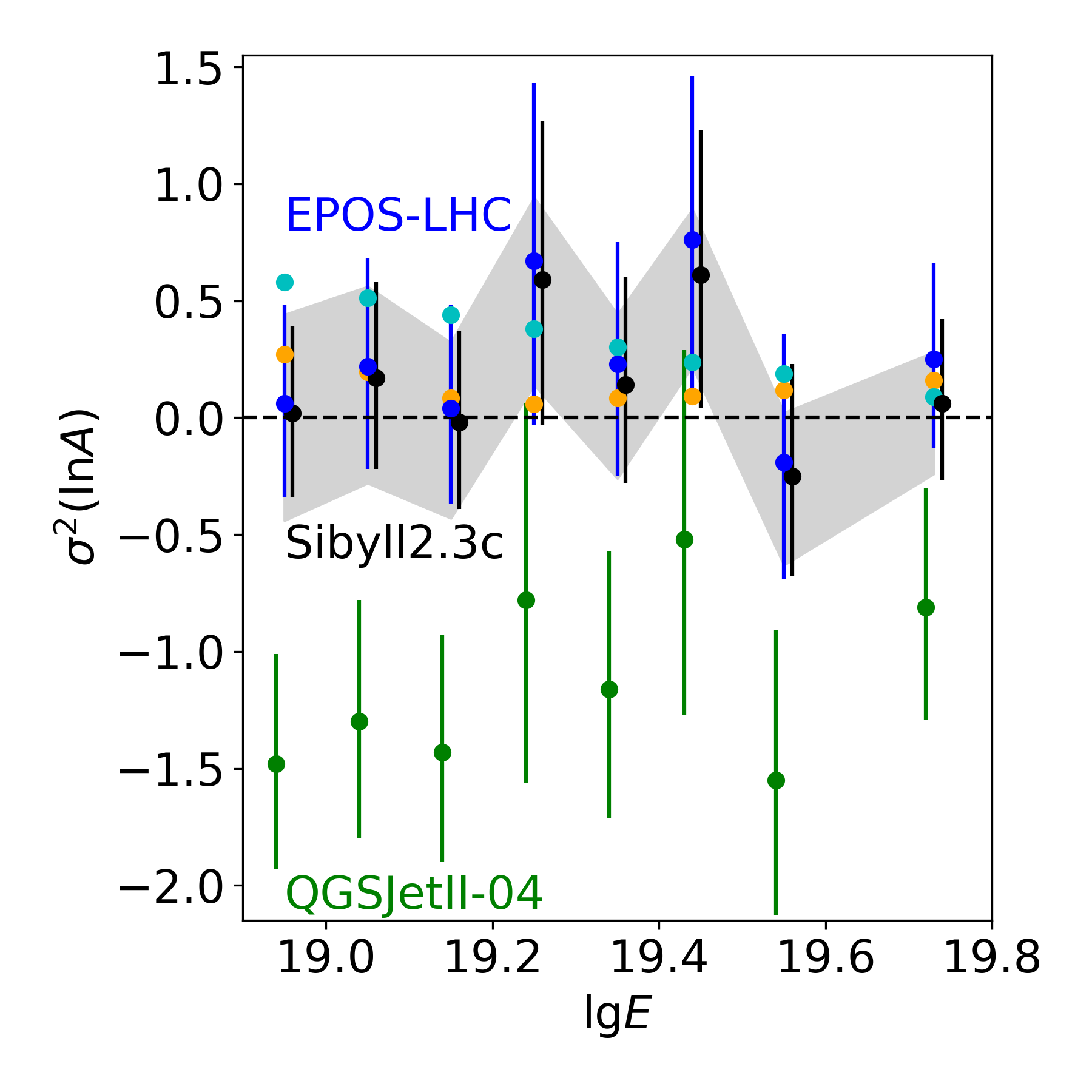}
\includegraphics[width=0.47\linewidth]{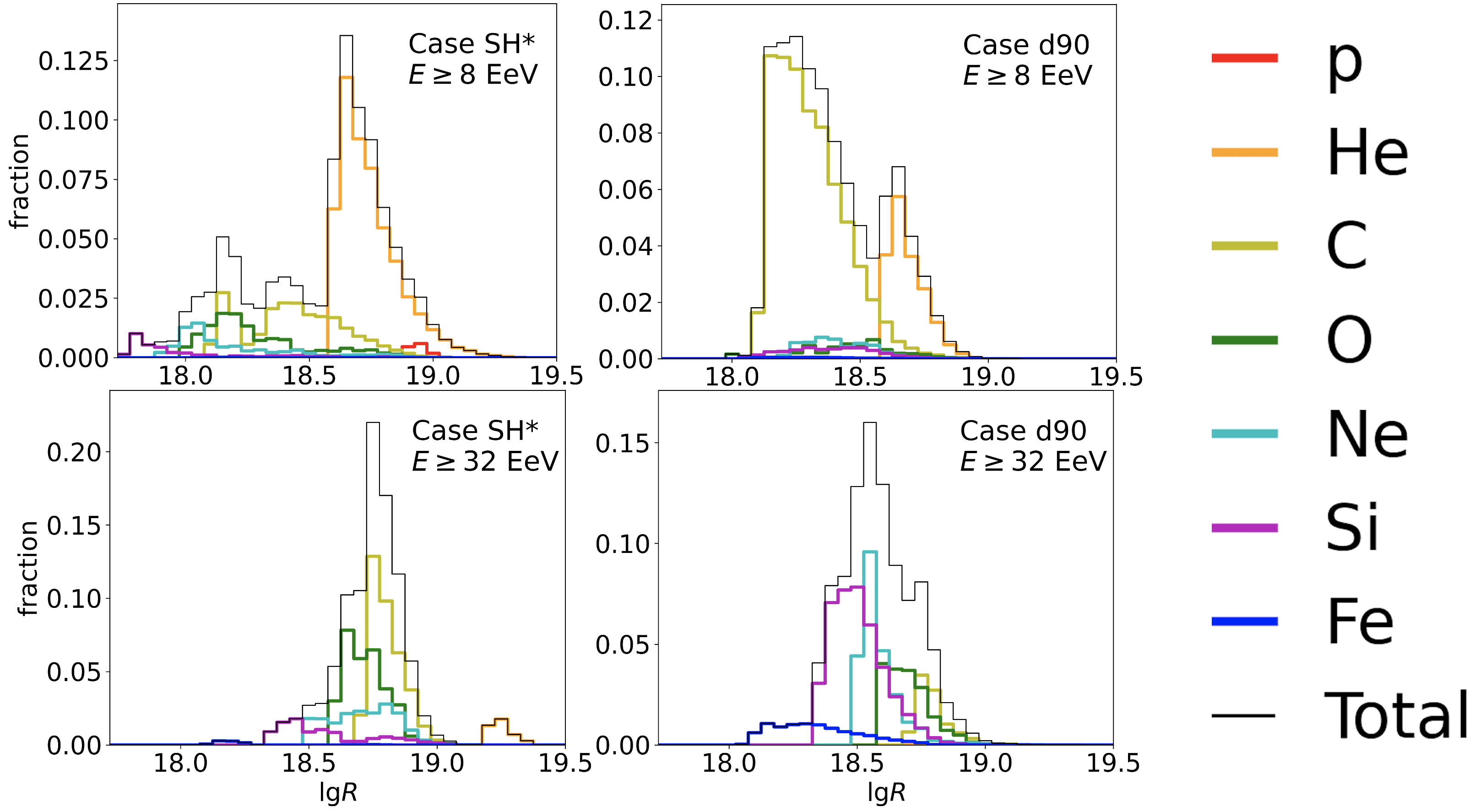}
\caption{Composition observables from shower development data.  
{ Left and middle panels:} The evolution with energy of $\langle\ln A\rangle$ and $\sigma^2(\ln A)$ with the statistical error, inferred using Sibyll2.3c (black), EPOS-LHC (blue) and QGSJETII-04 (green) from  \citet{ICRC2019_Auger_comp}.  The shaded bands mark the systematic error for Sibyll2.3c; the other two interaction models have similar systematic error. The cyan dots show our overall best fit to the dipole, hotspot and composition for the sharp-cutoff family of modeling, SH*, while the orange dots are the best fit for exponential attenuation modeling, with just dipole and composition being fit.  { Right panels:} The corresponding rigidity spectrum for the data set $\geq 8$ EeV (top) and 32 EeV (bottom) for the SH* (left) and d90 (right) analyses.}
\label{fig:compo_and_rigidity_spectrum}
\end{figure*}

\subsection{Mass composition}
The mass composition of UHE cosmic rays can be inferred from measurements of the longitudinal development of UHE air showers, with the use of hadronic interaction models. 
Figure~\ref{fig:compo_and_rigidity_spectrum} shows the mean and variance of $\ln A$ and their uncertainties, in eight energy bins as inferred by the Pierre Auger Collaboration \citep{ICRC2019_Auger_comp}
using three different HIMs: Sibyll2.3c, EPOS-LHC, QGSJETII-04.\footnote{It should be noted that QGSJETII-04 gives an unphysical $\sigma^2(\ln A)$.} We parameterize the evolution of the composition with energy as follows (abbreviating ${\rm log}_{10}(E) \equiv \lg E$): \\
\hspace*{0.05\textwidth}$\bullet \quad \langle\ln A\rangle=\alpha \lg E + \langle\ln A\rangle_{8\rm\, EeV}$ \\
\hspace*{0.05\textwidth}$\bullet \quad \sigma^2(\ln A)=\beta \lg E +{\sigma^2(\ln A)}_{8\rm\, EeV}$. \\
Thus the model composition is characterized by four parameters $\boldsymbol{\Omega}\equiv  \{\alpha, \beta,  \langle\ln A\rangle_{8\rm\, EeV}, {\sigma^2(\ln A)}_{8\rm\, EeV}\}$\footnote{For the purpose of reporting results, we quote a different $\boldsymbol{\Omega}\equiv[\langle\ln A\rangle_{\rm8-10\,EeV},\langle\ln A\rangle_{\geq 40 {\rm\,EeV}}, \sigma^2(\ln A)_{\rm8-10\,EeV},\sigma^2(\ln A)_{\geq 40 {\rm\,EeV}}]$ because it corresponds to $X_{\max}$ measurements (the first and last data points in Fig.~\ref{fig:compo_and_rigidity_spectrum} left and middle panels). The two $\boldsymbol{\Omega}$s are equivalent and can be used interchangeably.}. We consider seven chemical elements, p, He, C, O, Ne, Si, Fe, to reduce unphysical discontinuities for low A, and mix them with appropriate proportions to achieve the $\langle\ln A\rangle$ and $\sigma^2(\ln A)$ specified by a given choice of $\boldsymbol{\Omega}$ (see Appendix \ref{Appendix:fitting} for details). The relative abundances of the seven chemical elements in the $\geq 8$ EeV and $\geq 32$ EeV bins as a function of their rigidity ($R=E/Z$) are displayed in the right panel of Fig.~\ref{fig:compo_and_rigidity_spectrum}, for the two overall best-fitting SH* and d90 models.

\subsection{Fitting model parameters}
In this analysis we fit the parameters of the model described above to the following observations, or subsets of them:\\ 
{\it i)} Nine dipole components $d_{x},d_{y},d_{z}$ reported by Auger in the three energy bins: 8-16, 16-32 and $\geq 32$ EeV \citep{Auger_all_bins_2020}, denoted ``dipole" below;\\ 
{\it ii)} The observed arrival directions of 1288 events above 38 EeV observed by Auger, reconstructed from the Li-Ma sky map in \citet{ICRC2019_Auger_hotspot}, denoted ``events" below;\\
{\it iii)}  The $\langle\ln A\rangle$ and $\sigma^2(\ln A)$ inferred by Auger from $X_{\max}$ measurements in the 8 energy levels $\geq 8$ EeV, for each HIM \citep{ICRC2019_Auger_comp}.

The quality of the fit is given by the likelihood of obtaining the given measurements if the cosmic-ray intensity maps and composition are given by the LSS model, for any specified parameter set. For instance when fitting all three sets of data listed above, the likelihood function is
\begin{eqnarray}
\ln L \equiv \ln L(\mathrm{dipole}\mid\boldsymbol{\Theta};\mathrm{source}) + \ln L(\mathrm{events}\mid\boldsymbol{\Theta};\mathrm{source})\nonumber\\ + \ln L(\langle\ln A\rangle\mid\boldsymbol{\Omega};\mathrm{HIM}) + \ln L(\sigma^2(\ln A)\mid\boldsymbol{\Omega};\mathrm{HIM})\,,
\label{eq:likelihood}
\end{eqnarray}
where ``source" is the specified source distribution model (LSS or isotropic, and treatment of attenuation), ``HIM" refers to the hadronic interaction model used to infer the composition, and $\boldsymbol{\Theta}$ is the set of six parameters to optimize the fit: two parameters for the EGMF and GMF ($D_{\rm EG,5EV}$ and $\lambda_{\rm G}$), and the four composition parameters denoted by $\boldsymbol{\Omega}$.  

\begin{table*}
\centering
\begin{footnotesize}
\begin{tabular}{|c|c|c||c|c|c||c|c|c|}
\hline
Cases & & Iso & SH* & SH$_{\rm E}$* & SH & d90 & d90* & PP \\ \hline
\multirow{3}{*}{\rotatebox{90}{Model}} & Source & Isotropic & LSS & LSS & LSS & LSS & LSS & LSS \\ 
& Distance weighting & --- & Sharp & Sharp & Sharp & Exponential & Exponential & Exponential \\
& HIM & Sibyll2.3c & Sibyll2.3c & EPOS-LHC & Sibyll2.3c & Sibyll2.3c & Sibyll2.3c & ---\\ \hline
\multirow{6}{*}{\rotatebox{90}{Likelihood}} & $\ln L(\mathrm{dipole}\mid\boldsymbol{\Theta};\mathrm{source})$ & -3.4 & 14.5 & 13.8 & 20.6 & 13.4 & 10.2 & -6.0 \\
& $\ln L(\mathrm{events}\mid\boldsymbol{\Theta};\mathrm{source})$ & 0 (Ref) & 11.1 & 10.9 & 2.4${}^{\dagger}$ & 0.1${}^{\dagger}$ & 4.4 & 0.6${}^{\dagger}$  \\
& $\ln L(\langle\ln A\rangle\mid\boldsymbol{\Omega};\mathrm{HIM})$ & 4.4 &  4.0 & 3.8 & 1.8 & 4.1 & 3.5 & --- \\
& $\ln L(\sigma^2(\ln A)\mid\boldsymbol{\Omega};\mathrm{HIM})$ & -2.8 & -3.3  & -3.7 & -3.0 & -3.0 & -3.9 & --- \\ \cline{2-9}
& Sum of three $\ln L$s without $\ln L(\mathrm{events})$ & -1.9 & 15.3 & 13.8 & 19.5 & 14.4 & 9.8 & --- \\
& Sum of four $\ln L$s (Eq.~\eqref{eq:likelihood}) & -1.9 & 26.3 & 24.7 & 21.8 & 14.5 & 14.2 & --- \\ \hline
\multirow{7}{*}{\rotatebox{90}{Best-fit parameters}} 
& $\lg D_{\rm EG,5EV}$ & --- & $2.79^{+0.60}_{-0.20}$  & $2.79^{+0.60}_{-0.20}$ & $2.39^{+0.40}_{-0.40}$ & $\infty$ & $\infty$ & $\infty$ \\
& $\lg\lambda_{\rm G}$ & --- & $1.58^{+0.10}_{-0.08}$ & $1.58^{+0.13}_{-0.08}$ & $1.82^{+0.13}_{-0.16}$ & 1.95 & 1.66 & 1.48 \\
& $\langle\ln A\rangle_{\rm8-10\,EeV} $  & 2.02 & $1.76^{+0.19}_{-0.15}$ & $1.64^{+0.18}_{-0.16}$ & $2.11^{+0.36}_{-0.29}$ & 2.12 & 2.27 & 0 \\
& $\langle\ln A\rangle_{\geq 40 {\rm\,EeV}} $ & 3.19 & $2.87^{+0.17}_{-0.10}$ & $2.84^{+0.13}_{-0.08}$ & $3.27^{+0.33}_{-0.30}$ & 3.29 & 3.44 & 0 \\
& $\sigma^2(\ln A)_{\rm8-10\,EeV} $ & 0.31 & $0.48^{+0.27}_{-0.20}$ & $0.47^{+0.27}_{-0.24}$ & $0.40^{+0.30}_{-0.18}$ & 0.27 & 0.23 & 0 \\
& $\sigma^2(\ln A)_{\geq 40 {\rm\,EeV}}$ & 0.19 & $0.09^{+0.32}_{-0.05}$ & $0.09^{+0.34}_{-0.05}$ & $0.28^{+0.28}_{-0.19}$ & 0.16 & 0.62 & 0 \\
\cline{2-9}
& $B_{\rm EG}$ if $\lambda_{\rm EG}=0.2$ Mpc  & --- & $0.32^{+0.08}_{-0.16}$ & $0.32^{+0.08}_{-0.16}$ & $0.50^{+0.13}_{-0.25}$ & 0 & 0 & 0 \\
\hline
\multirow{2}{*}{\rotatebox{90}{hotspot}} 
&\hspace*{-0.2in}\makecell{Number of events in $27^\circ$ circle\\centered at ($309.7^\circ,17.4^\circ$).  Obs=188} & $125^{+11}_{-11}$  & $154^{+12}_{-11}$ & $156^{+12}_{-12}$ & $135^{+11}_{-11}$ & $139^{+11}_{-11}$ & $138^{+11}_{-11}$ &  $160^{+12}_{-12}$ \\
\cline{2-9}
&\hspace*{-0.2in}\makecell{Li-Ma significance in $27^\circ$ circle\\centered at ($309.7^\circ,17.4^\circ$). Obs=5.6}& $0.0^{+1.0}_{-1.0}$ & $2.7^{+1.1}_{-1.0}$ & $2.8^{+1.1}_{-1.1}$ & $0.9^{+1.0}_{-1.0}$ & $1.3^{+1.0}_{-1.0}$ & $1.2^{+1.0}_{-1.0}$ & $3.2^{+1.0}_{-1.1}$ \\
\hline
\end{tabular}
\end{footnotesize}
\caption{\small Summary of model parameters and results.  In cases SH*, SH$_{\rm E}$* and SH, the sharp-cutoff horizon is used, while for the remainder of examples the exponential attenuation weighting of Eq.~\eqref{eq:d90} is used.  
The best-fit parameters are shown as median with $1\sigma$ confidence levels (i.e. 16\% and 84\% percentiles).  The SH cases allow for an EGMF, while the EGMF is set to zero for the rest. For SH and d90, only the dipole and composition are fit, while for the cases with * in their names, the events above 38 EeV are also included in the fit.  For the pure proton case, ``PP'', only the dipole (not composition nor events) is fit. Note that the total $\ln L$ should only be compared between cases where the same datasets are being fit.  Entries in the $\ln L$ section marked with a dagger (${}^{\dagger}$) indicate the corresponding data were {\it not} included in the fitting for that case.
In the bottom two rows, the hotspot results are calculated from millions of mock data sets generated from the model arrival map above 38 EeV with the best-fit parameters. The confidence level represents the statistical uncertainty in the mock data datasets, and does not represent the uncertainty due to the uncertainty in the best-fit parameters.}
\label{tab:likelihood}
\end{table*}

\section{How good is the LSS ansatz?}

We have performed various studies of the LSS ansatz, using the framework described in the previous section and fitting to more or less data. The results for some instructive cases are reported in 
Table~\ref{tab:likelihood}.  The column  ``Iso"  shows the analysis applied to an isotropic skymap weighted by the Auger exposure, while subsequent columns are for the LSS source hypothesis, exploring different treatments of the attenuation (sharp cutoff for cases labeled SH and exponential for the rest) and composition (fitting to Sibyll2.3c or EPOS-LHC, or simply pure proton (PP)). Cases labeled with an asterisk denote that the parameters of the model have been fit to the events above 38 EeV and not just the nine dipole components.  A discussion of the lessons learned is given below.

The six ``Likelihood" rows of Table~\ref{tab:likelihood} give the log-likelihood for each case, broken down by datasets and the total. In each case, the lnL values for datasets not included in the fit are marked by a dagger. Note that the larger $\ln L$, the better the fit.  The next six rows give the best-fit parameters and their $68\%$ confidence level range, and in the following row we convert the EGMF constraint on $D_{\rm EG,5EV}$ into more familiar terms by quoting the $B_{\rm EG}$ derived from $D_{\rm EG,5EV}$ taking $ \lambda_{\rm EG}=0.2$~Mpc. The final rows give ``predictions" for the Auger hotspot in each case.  Note that cases SH* and d90* include the events above 38 EeV as a constraint, so they naturally give better fits to the data above 38 EeV than do the other cases.  


At various points in this paper we have adopted models d90 and SH* to illustrate different points, because d90 or d90* use the most realistic treatment of attenuation available now, and SH* gives the overall best-fit.  We select d90 rather than d90* because only fitting the dipole is most conservative, given that the events above 38 EeV may get a significant contribution from individual sources -- moreover the fit quality of d90 and d90* are similar.  Conclusions that emerge independently of the horizon treatment (sharp cutoff vs. exponential) can be taken to be robust and likely to endure in a more comprehensive study.  We discuss below what these conclusions are, and which questions need a more refined analysis to decide.  


\subsection{Extragalactic magnetic field}
So far we have only explored the impact of EGMF in the SH approach.  There, the inferred EGMF is very small ($0.32^{+0.08}_{-0.16}$ nG or $0.50^{+0.13}_{-0.25}$ nG) depending on whether the model is fit to all of the data including the events above 38 EeV (SH*), or not (SH).  Either value is compatible with typical expectations in the homogeneous EGMF approximation, and implies a lower impact of EGMF on the anisotropy than would obtain for the 10~nG used for the examples in \citep{GAP08}.  
We remark in passing that locally strong fields such as observed in massive galaxy clusters will concentrate UHECRs near their sources -- effectively increasing the apparent source contrast.  Near-future studies of source ``bias" with respect to the (predominantly Dark) matter distribution will probably find the two effects quite degenerate.

\subsection{Sensitivity to HIM}
Cases SH* and SH$_{\rm E}$* are the same, except that the first case fits to the Auger composition inferred taking Sibyll2.3c as the HIM, and the second one uses the EPOS-LHC composition.   The two cases are nearly identical in fit quality and inferred parameters;  we adopt Sibyll2.3 below since it is slightly preferred.  Fig.~\ref{fig:compo_and_rigidity_spectrum} shows that the best-fitting composition parameterizations found for cases d90 and SH* are quite consistent with the Auger composition results.

\begin{figure*}
\centering
    \centering
         \centering
         \includegraphics[width=0.28\linewidth]{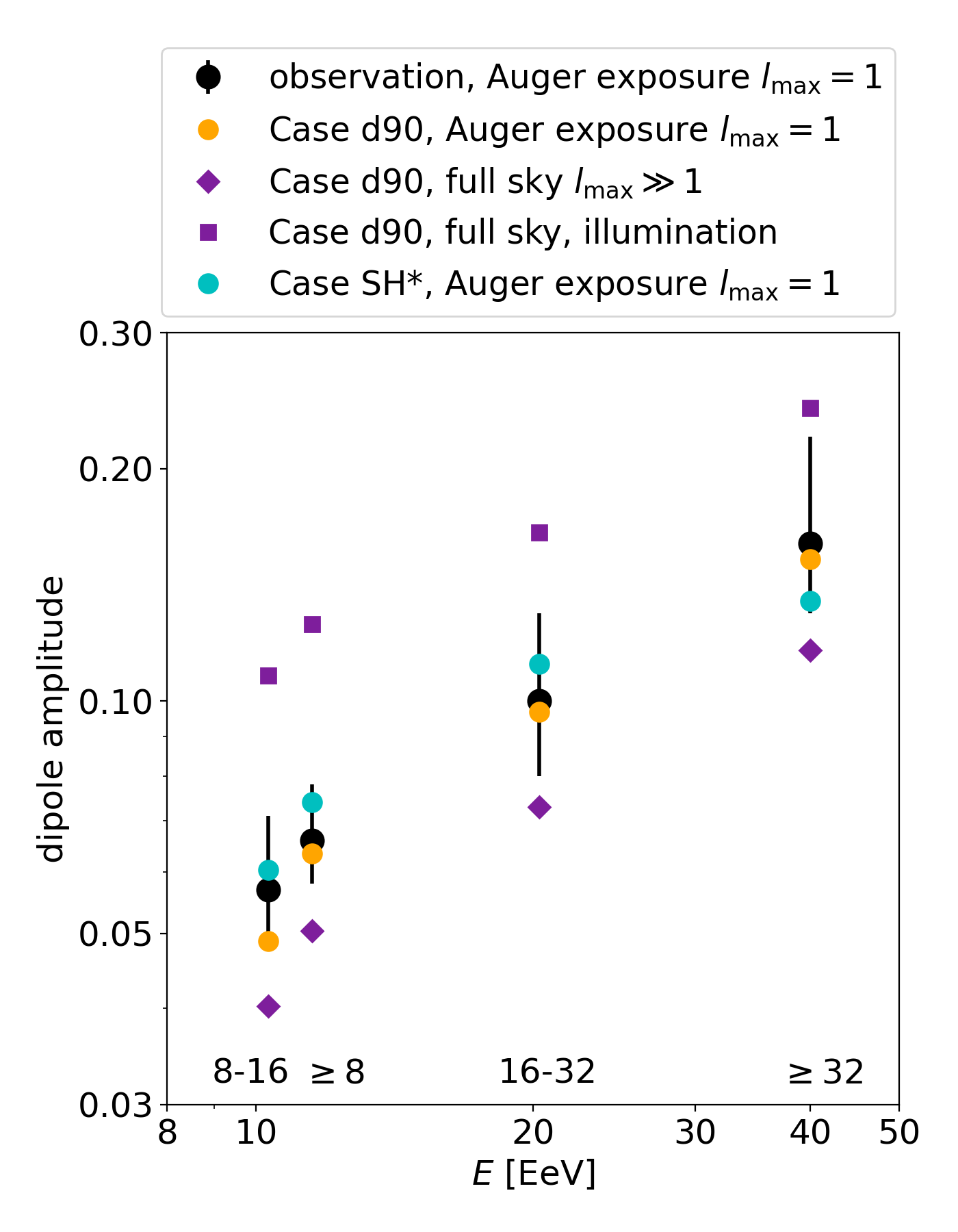}
         \includegraphics[width=0.69\linewidth]{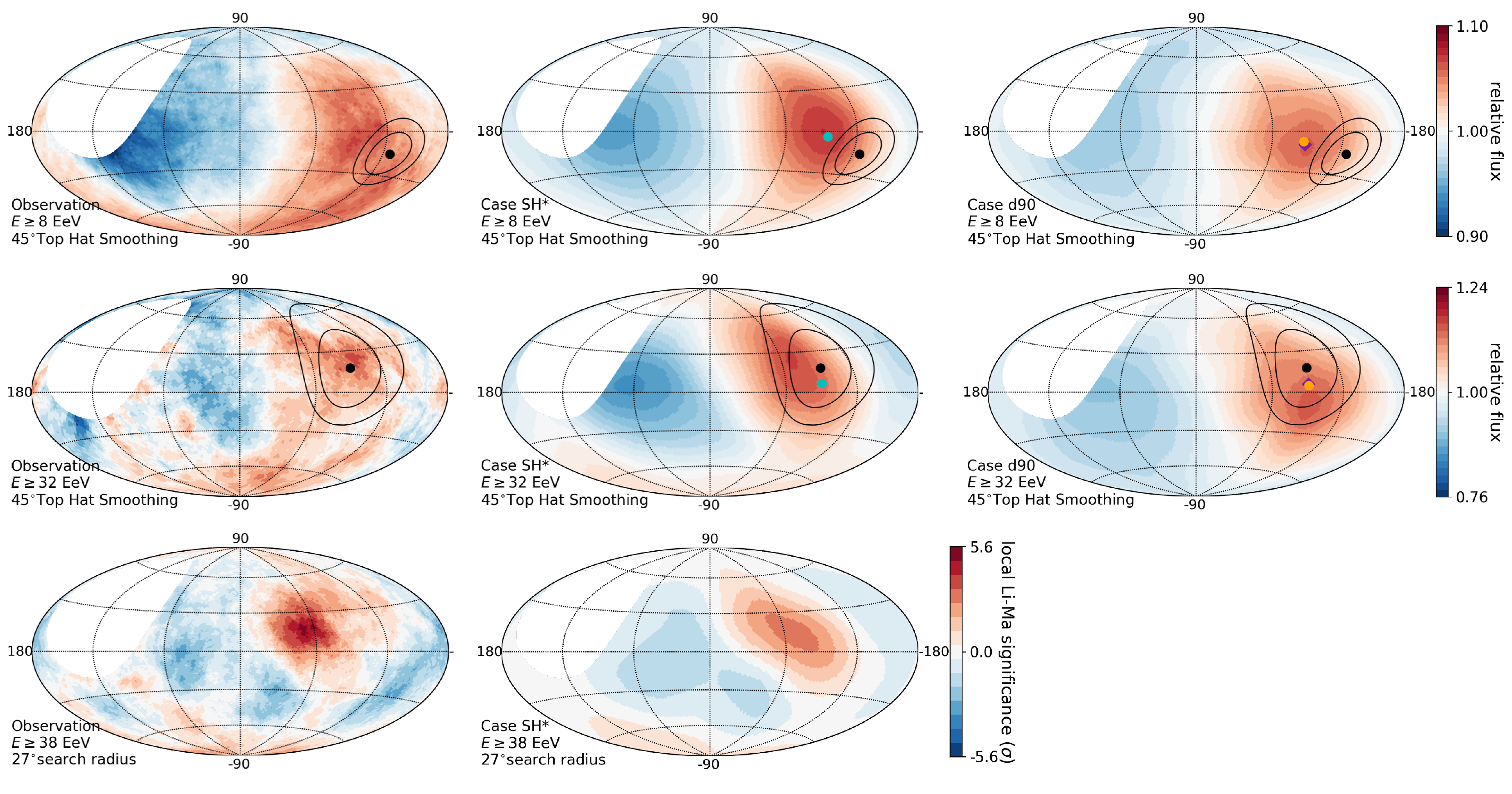}
\caption{Comparison between observed and model anisotropy.   {\it In the left panel:} The magnitude of the LSS-induced dipole amplitude for different energy bins is compared to the data (black dots with error bars).  The orange and cyan dots show the prediction for the d90 and SH* models given Auger's exposure.  For comparison, the purple diamonds show the d90 prediction with full-sky exposure and the purple squares show the prediction without GMF processing; see Table~\ref{tab:result_above8} in the Appendix for the full set of dipole components. {\it Skymaps:} The left column of sky maps show the Auger data maps from \citet{Auger_Science_2017,Auger_4bin_2018} and \citet{ICRC2019_Auger_hotspot}. The top hat maps from \citet{Auger_Science_2017,Auger_4bin_2018} are replotted using the same smoothing method as for the model maps as detailed in Sec.~\ref{sec:top_hat_method}. The closed black curves show the 68\% and 95\% confidence level regions of dipole direction, with the black dot indicating the center. Middle (right) columns: SH* (d90) predictions; note that sampled with the same number of events as in the data maps, the predicted LSS maps are similarly blotchy as the data maps.  Top (middle) rows:  $\geq 8$ EeV ($\geq 32$ EeV), smoothed by a $45^\circ$ top-hat for comparison to the Auger sky maps. The colored dot is the dipole direction of the model, reconstructed with detector exposure and $l_{\rm max}=1$; the purple diamond is the dipole direction for full-sky exposure and $l_{\rm max}\gg1$. Bottom row: Li-Ma significance above 38 EeV with search radius of 27 degrees, for Auger (left) and SH* model (middle).  
}
\label{fig:dip_AM}
\end{figure*}

\subsection{Dipole Anisotropy Predictions}
\label{sec:dipole_anisotropy}

Figure \ref{fig:dip_AM} summarizes the LSS model predictions for the dipole anisotropy and its energy dependence. The left panel of Fig.~\ref{fig:dip_AM} shows that the observed magnitude of the dipole amplitude and its energy evolution can be well-described by the LSS model. 
As can be seen from Table 1, the overall quality of fit is marginally better for SH* than for the d90 model, and better still for SH which does not fit to the events.  The best-fitting composition changes from SH* to d90, but without changing the quality of fit to the composition data. In principle the exponential attenuation treatment should be more accurate than the sharp cutoff models but, as noted above, the source spectral index of -2.4 underlying the $d_{90}(A,E)$ values may not be the optimal representation of the spectrum, and the energy-averaged character of the present treatment (Eq.~\eqref{eq:d90}) is only a first approximation.  

The right panels of Fig.~\ref{fig:dip_AM} show the observed sky maps compared to the SH* and d90 model predictions for the arrival maps, for the energy bins $\geq 8$ EeV and $\geq  32$ EeV.  (Also the LiMa map above 38 EeV, discussed below.)    Unlike the observation sky map $\geq 8$ EeV, the sky above $ 32$ EeV does not look like a simple dipole. It has two distinct regions of excess, one in the Northern sky and another toward the Galactic South Pole. The LSS model gives a reasonable explanation for this. The increase of energy threshold results in the decrease of horizon size, which in turn leads to substantial increase in the percentage of flux from the Local (Virgo) and Hydra-Centaurus superclusters which give excess to the northern sky.  It also increases the percentage flux from Perseus-Pisces Supercluster which, after processing by the GMF, gives regions of excess as seen in the Southern hemisphere.  
These effects can be seen in  Fig.~\ref{fig:3shells}.  The pattern of excess seen in Fig.~\ref{fig:dip_AM} is evocative of the pattern from Perseus-Pisces (after GMF deflections) and suggests that an improved treatment of attenuation could lead to a somewhat stronger contribution from the Perseus-Pisces region and enable a better fit the Auger observations $\geq 32$~EeV.

\subsection{Dipole Anisotropy, Discussion}

As already mentioned, in the LSS model the observed UHECR dipole anisotropy is the result of an interplay between composition-dependent horizon and rigidity-dependent deflection/diffusion in the GMF. Energy-loss processes should be well-determined from laboratory measurements   \citep[but see][]{batista+2015prop}, so if the matter distribution is sufficiently well-known, anisotropies can contribute to disentangling composition and GMF since they impact the arrival direction maps differently. Some sensitivity to EGMF diffusion may also be possible, as discussed earlier.

The left column of Fig.~\ref{fig:3shells} shows the surface density contrast of individual 40 Mpc shells out to 120 Mpc. 
In the absence of EGMF diffusion, the illumination maps (shown in Fig.~\ref{fig:figure_1} and Appendix \ref{Appendix:allshells}) are superpositions of the surface density of 1 Mpc shells depending on attenuation and composition. All the shells out to 360 Mpc are shown in Appendix \ref{Appendix:allshells}.   
One sees, as expected, that more distant shells which average over a larger volume show less contrast. Also shown in Fig.~\ref{fig:3shells} are the arrival direction maps corresponding to the given shells, after deflection and diffusion in the GMF, 
as predicted by the d90 and SH* models for the $E\geq 8$ dataset .  Shells for more models, extending to larger distances and including the $\geq 32$ EeV predictions are shown in Appendix Figs.~\ref{fig:allshells_SHasterisk}--~\ref{fig:allshells_PP}.
For a given model, the arrival maps for the different energy thresholds have quite a similar pattern, due to the fact that the peak of the rigidity distributions changes rather little with threshold, as seen in the left panel of Fig.~\ref{fig:compo_and_rigidity_spectrum} and as expected for magnetic acceleration and energy losses preserving $E/A$, except for protons. Thus, the dominant reason that the direction and character of UHECR anisotropies change with energy is the horizon effect.  In addition, as energy increases and the horizon is reduced, stochastic ``cosmic variance" effects become more important since the number of sources being sampled is reduced.

Setting aside this cosmic variance for our initial study, let us look more closely at how the relative contribution of flux from different distances in the nearby universe depends on energy, shown in Fig.~\ref{fig:figure_1} and Appendix Fig.~\ref{fig:horizon_flux}. For example in the sharp-cutoff approximation, the Local Supercluster contributes 8\% of the total UHECR flux for 8-16 EeV and 33\% above 32 EeV; the corresponding numbers of the d90 model are 19\% and 48\%. The difference between attenuation treatments emphasizes the need to correctly model the attenuation, but the commonality underscores two main consequences of the LSS model for the anisotropy: the dipole amplitude increases with energy and the anisotropy pattern changes with energy. 


The composition affects the anisotropy in two distinct ways: \\ 
{\it i)}{ \it GMF deflections and dispersion are rigidity dependent and generally increase as rigidity drops.}  \\
The d90 model favors a heavier composition than the SH* model, and as a result, its arrival maps look quite different. In Appendix \ref{Appendix:allshells}, we test a special model ``d90sp'' that uses the same exponential attenuation as d90, but uses the same composition as SH*. The comparison of illumination maps and arrival maps between d90 and d90sp illustrates the composition effect. A more extreme example is the pure proton model shown in Appendix Fig.~\ref{fig:allshells_PP}, for which the deflection provided by the GMF is far too small to be consistent with the observed dipole. Note that a possible EGMF would only affect the amplitude and not improve the direction of the dipole, so our not allowing for an EGMF does not limit the validity of this conclusion.

The impact of energy threshold on GMF deflections for a single model can be seen in comparing the middle and right columns of the entire set of figures \ref{fig:allshells_SHasterisk}-\ref{fig:allshells_PP}.  Possibly nonintuitively, the main impact is on the degree of contrast, because to first approximation the mean rigidity which determines {\it where} the structures are, does not change with rigidity threshold.  However the spread in rigidities and hence degree of smearing, increases at lower energies due to the larger jumps between $A$ values. \\
{\it ii)} {\it The weight of different shells at a given energy is different for each $A$, due to the $A$-dependence of energy-loss-lengths.}  We assay the importance of this effect by comparing the illumination and arrival maps of d90 and d90sp models in Figs.~\ref{fig:allshells_d90} and \ref{fig:allshells_d90sp}.

\subsection{Intermediate Scale Anisotropies} \label{sec:hotspot}

As energy goes still higher, the separation of two regions of excess becomes even more noticeable. 
The arrival map above 38 EeV observed by Auger can be represented by a hotspot centered at ($309.7^\circ,17.4^\circ$) in Galactic coordinates and an excess around the Galactic South Pole. Auger found the hotspot has highest Li-Ma  significance \citep{Li-Ma} with a $27^{\circ}$ angular radius. The local Li-Ma significance is 5.6$\sigma$ and the post-trial significance is 3.9$\sigma$ \citep{ICRC2019_Auger_hotspot}.  The last column of Fig.~\ref{fig:dip_AM} shows the LiMa map of events above 38 EeV from Auger \citep{ICRC2019_Auger_hotspot} and the predicted LiMa map from our case-study SH*.

SH* is fit to both the dipole and the events above 38 EeV, to test whether the LSS alone, without a single source or sources, can give rise to the observed structure as was hinted by \citet{GPH18}.
To check the compatibility of an LSS model with hotspot observations, we performed a hypothesis test using the same test statistic as in \citet{Auger_starburst_gAGN,TA_test_starburst} based on the log-likelihood ratio. The methodology is discussed in Appendix \ref{Appendix:fitting}. The test statistic $\mathrm{TS} = 2 \ln L(\mathrm{events}\mid\boldsymbol{\Theta};\mathrm{source})\approx 22$. By simulating mock data sets, we find that the isotropic model is disfavored against the SH* model by 4.8 $\sigma$ (see Appendix~\ref{Appendix:lima}). 

The strength of the predicted LSS hotspot for our best-fitting description, case SH*, is weaker than the one observed. This could be the result of fitting to the arrival directions of all events above 38 EeV, which contain more information than just the strength of the hotspot, including the shape of the hotspot and the South Pole excess. 
To see whether a variation of the model parameters can give a similarly significant hotspot as observed, we carried out the exercise reported in the Appendix Table \ref{tab:more_cases} entry named Case SH(better hotspot)*, for which the objective function also includes the number of events inside the hotspot. The conclusion is that indeed the hotspot can be readily described, if it is included in the fitting, without significant damage to the fit to dipole components and events above 38 EeV. 

In the LSS model, both the hotspot and the south pole excess may be reflections of the large-scale structure of the Universe. In particular, for the SH* model $\sim 40\%$ and $\sim 25\%$ of cosmic rays inside the hotspot come from the Local and Hydra-Centaurus superclusters respectively and these naturally lead to the observed structures after deflection in the GMF.   It is important to note that the LSS models do not include specific galaxies, such as Cen A/NGC4945/NGC253.  As shown in  \citep{Auger_starburst_gAGN} individual classes of sources may instead be the origin of the hotspot and other structure seen above 38 EeV, if net GMF deflections are small enough. Auger also performed a correlation study with 2MRS galaxies, as a proxy for local large scale structure, however that study cannot be compared to the present one because  deflections in the GMF are ignored and 2MRS is a flux-limited catalog with a high flux threshold (K-band magnitude 11.75), hence subject to greater cosmic-variance issues than {\it CosmicFlows-2}. 

We note that -- independently of attenuation treatment and composition -- no prominent structure in the direction of the TA hotspot is expected 
from the LSS.  This can be seen in the all-sky prediction plots of Figs.~\ref{fig:allshells_SHasterisk}-~\ref{fig:allshells_PP}.  If the TA hotspot holds up as statistics increase, that would be strong evidence for an individual source not captured by the LSS treatment.

\subsection{Composition}\label{sec:compo}

Conservatively applying the LSS analysis only to the dipole data, one sees from Table 1 cases d90 and SH that roughly comparable, acceptable fits can be obtained for the different attenuation treatments -- with the resulting best-fit compositions adjusting to give compatibility.  This means that with good composition information -- after HIM models are improved and vetted with future air-shower-development and accelerator observations -- we can test in detail the attenuation modeling.  Conversely, when more accurate attenuation modeling is in hand, LSS modeling of the dipole can constrain composition.  The intermediate scale anisotropies at higher energy are even more sensitive to composition and can provide important additional information -- with self-consistency probably making it clear whether the Auger hotspot can be explained without a prominent individual source.

\section{Summary and Conclusions}

The postulate that the UHECR source distribution follows the matter distribution of the local Universe, taking into account deflections in the GMF and possible diffusion in the EGMF, provides quite a good accounting of the magnitude, direction and energy dependence of UHECR anisotropies $\geq 8$ EeV. 

A main feature of this LSS hypothesis is the horizon effect, a reduction of the observable UHECR universe with energy that leads to an increase of the contribution of nearby superclusters at the highest energies, whose contributions provide distinctive signatures in the UHECR sky.  The current implementation of the LSS model can account for the evolution of the observed dipole amplitude with energy but, as can be seen in Fig.~\ref{fig:dip_AM}, the LSS dipole direction for the $\geq 8$~EeV bin is off by some degrees from the observed direction (95\% confidence level circle). This can be due to our still-approximate treatment of the horizon effects, our incomplete knowledge of the local matter distribution and/or our incomplete knowledge of the cosmic and Galactic magnetic fields.  A refined GMF model is under development \citep{UF_progress_2019}, and our knowledge of the LSS is evolving with new discoveries such as the South Pole Wall \citep{pomarede2020cosmicflows} that are only partially taken into account in CosmicFlow-2, so we can reasonably expect better fidelity in the future.




Conclusions that can already be reached include the following.\\ \\
    $\bullet$ The direction of the large scale UHECR dipole and its evolution with energy appear to be the result of the interplay between the local large scale matter distribution -- especially the orientation of the Galaxy relative to the Super-Galactic Plane -- and deflection in the large-scale coherent poloidal (X-shaped) halo field of the Milky Way.  Improvements in our understanding of the Galactic magnetic field can shift the predicted direction of the dipole some degrees, while the strength and orientation of the anisotropy arriving to the Galaxy is mainly sensitive to the horizon and LSS distribution.  Both effects are sensitive to the composition.\\ \\
     $\bullet$  A pure proton composition can be excluded, unless the dipole anisotropy is not due to large scale structure.  \\ \\
     $\bullet$  Both the Auger hotspot and the excess near the South Galactic Pole seen above 38 EeV can potentially be due to the large scale distribution of matter, rather than individual dominant 
     sources. An interesting future work, after a more accurate treatment of the attenuation has been implemented, will be to quantitatively compare the quality of such a description to the starburst galaxy fit~\citep{Auger_starburst_gAGN}.\\ \\
     $\bullet$  UHECR source scenarios with many lower power sources rather than a few individual high-power sources, or slow acceleration in large scale accretion shocks, remain competitive.  The TA hotspot, currently at lower significance~\citep{ICRC2019_TA_hotspot}, seems to require an individual source if it is real; this, and constraints on the number density of sources, will be the topic of a separate report.\\ \\
     $\bullet$  In the future, with higher fidelity in the attenuation modeling, the observed anisotropies from the known large scale structure should provide a powerful constraint on the GMF, finally overcoming the chicken-and-egg problem of needing to know the sources in order to use deflections to improve the GMF so that deflection predictions can be made more reliable as needed to identify the sources.\\ \\

\section{Acknowledgments}
We thank Yehuda Hoffman for his permission to use the density field from \citep{Hoffman2018},  Daniel Pomar\`ede for the 3D visualization, Marco Muzio, Olivier Deligny, Denis Allard and Anatoli Fedynitch for valuable discussions, Shenglong Wang for assistance with the use of the NYU high performance computing facility, and members of the Pierre Auger Collaboration for helpful comments on the manuscript; GRF acknowledges membership in the Pierre Auger Collaboration. The research of NG was supported by New York University and the Simons Foundation, and that of GRF by NSF-1517319 and NSF-2013199. Some of the results in this paper have been derived using the healpy \citep{Healpy_Zonca2019} and HEALPix \citep{HEALPix} package.

\bibliography{main}
\onecolumngrid 
\appendix
\restartappendixnumbering

\section{Illumination map calculation procedure}\label{Appendix:illumination}
To calculate the illumination map, (i) we assume that the 3D cosmic-ray source distribution follows the LSS; (ii) we treat the UHECR diffusion in the EGMF by taking it to be homogeneous and characterized by a diffusion coefficient $D_{\rm EG}$.

\subsection{Source distribution}
We assume that the UHECR source density is proportional  to the matter density field, $\rho(r,\hat{e})$, where $r$ is the radial coordinate and $\hat{e}$ the direction in the sky. The density contrast is $\delta(r,\hat{e})=1-\rho(r,\hat{e})/\bar{\rho}$, where $\bar{\rho}$ is the mean density of the universe. 
To model the matter density, we use the same set of simulations as in \cite{GPH18}, i.e., cosmological simulations constrained by the {\it CosmicFlows-2} data \citep[CF-2,][]{tully2013cosmicflows}. Up to a distance of $350$ Mpc, this uses the quasi-linear density field  presented in \cite{Hoffman2018} for which the angular resolution of the boxes is~1.5-2 Mpc. (Note that this is larger than the extent of the 2MRS catalog which is flux-limited, whose sampling falls off beyond $\sim100$ Mpc.) Beyond 350 Mpc, 
the density field is obtained in the linear regime using a series of constrained realisations, within a $1830$ Mpc depth \citep[see][for more details]{GPH18}. Beyond $1830$ Mpc we simply assume that the Universe is homogeneous.  
Therefore, thanks to the CF-2 matter density field, we have the intensity sky maps of different shells of distances. In the "shell image", the intensity of each pixel is proportional to the source density in that pixel, $I_0(r,\hat{e}')\equiv\delta(r,\hat{e}')$, 
The angular resolution of each shell image is 1 degree. 


\subsection{Single source image}
The intensity profile of a cosmic-ray source depends on the scattering properties of the UHECRs in the EGMF.  The diffusion coefficient, $D_{\rm EG}$, 
can be approximated by a fitting function taking into account both the resonant and non-resonant diffusion regimes \citep{GAP08},
\begin{equation}
D_{\rm EG}\approx0.03\left(\frac{\lambda_{\rm Mpc}^2 E_{\rm EeV}}{ZB_{\rm nG}}\right)^{{1}/{3}}+0.5\left(\frac{E_{\rm EeV}}{ZB_{\rm nG}\lambda_{\rm Mpc}^{0.5}}\right)^{2} {\rm Mpc^2 \,Myr^{-1}}
\label{eq:dcoef}
\end{equation}
where $Z$ is the charge of the cosmic-ray,   $E_{\rm EeV}$ is its energy measured in EeV, 
$B_{\rm nG}$ the EGMF strength in nG and $\lambda_{\rm Mpc}$ the EGMF coherence length $\lambda_c$ in Mpc.

\begin{figure*}
\centering
\includegraphics[width=0.4\linewidth]{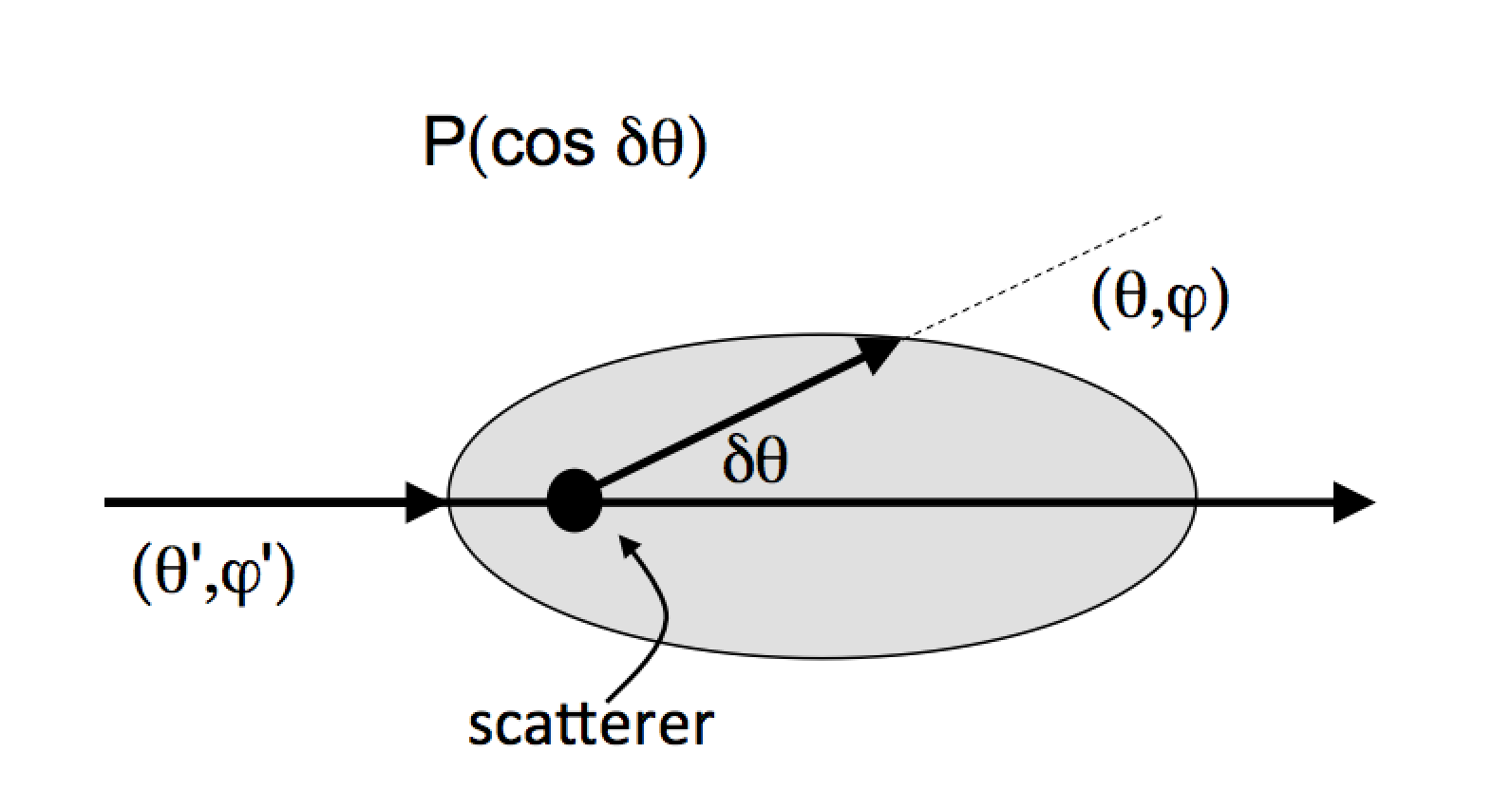}
\caption{The definition of the different quantities used in the  calculations and the single scattering angle $\cos\delta\theta$. The gray shaded area illustrates the phase function $P(\cos\delta\theta)$ (Eq.~\eqref{eq:phase_function}) introduced by \citet{Henyey:1941gpa}, which, by the variation of one parameter $q$, ranges from backscattering through isotropic scattering to forward scattering. Note that for cosmic rays, there is only isotropic (resonant scattering) and forward scattering (weak-scattering).}
\label{fig:phase_function}
\end{figure*}

The single scattering angle depend on the  optical depth, 
$\tau={rc}/{D_{\rm EG}}~$,
where $r$ is the distance from the source. It is characterized by the rms value 
\begin{equation}
\braket{\delta\theta^2}\sim(\lambda_c/ r_L)^{\kappa}{\braket{\delta B^2}/B^2},
\label{deltatheta}
\end{equation}
where $r_L$ is the Larmor radius $r_L = 1.1 \, ({\rm Mpc} E_{\rm EeV}/ (Z \, B_{nG})$) and  $\kappa=2$ for $\lambda_c\leq r_L$ and 
$\kappa=-2/3$ for $\lambda_c> r_L$ for Kolmogorov turbulence \citep{2008PhRvD..77l3003K}.   We assume strong turbulence ($\braket{\delta B^2}/B^2 =1$), so $\braket{\delta\theta^2}\sim(\lambda_c/ r_L)^{\kappa}$. 

To estimate the single source image (e.g. intensity sky map), we follow the same formalism as diffusion of light in scattering media,  described by the radiative transfer theory \citep{chandrasekhar1960}. In the case of a spherically symmetric atmosphere, the Radiative Transport Equation (RTE) which describes the change in flux through an infinitesimal volume is given by:
\begin{equation}
\mu\frac{\partial I}{\partial\tau}+\frac{1-\mu^2}{\tau}\frac{\partial I}{\partial\mu}
=-\frac{I(\tau,\mu)}{4\pi}\int_0^{2\pi}\int_{-1}^{1}P(\cos\delta\theta)I(\tau,\mu')d\mu'd\phi'
\label{eq:single_source_intensity}
\end{equation}
where $P(\cos\delta\theta)$ is the phase function, $\cos\delta\theta$ is the cosine of the angle between the incident direction ($\theta',\phi'$) and the scattered one in the direction ($\theta,\phi$), $\mu=\cos\theta$ and $\mu=\cos\theta'$ (see Fig.~\ref{fig:phase_function} for clarity).

\begin{figure*}
\centering
    \centering
         \centering         \includegraphics[width=0.3\linewidth]{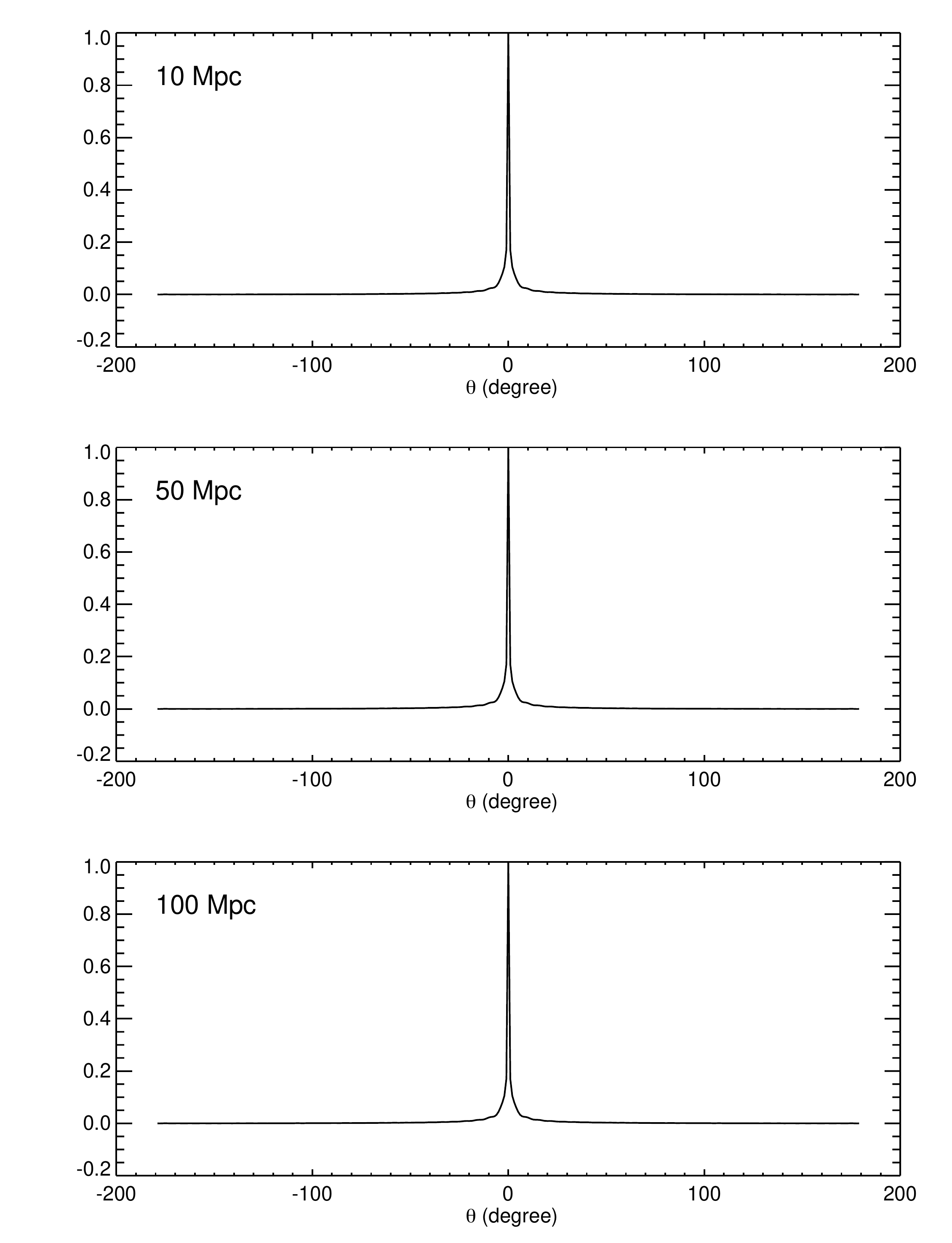}
        \includegraphics[width=0.3\linewidth]{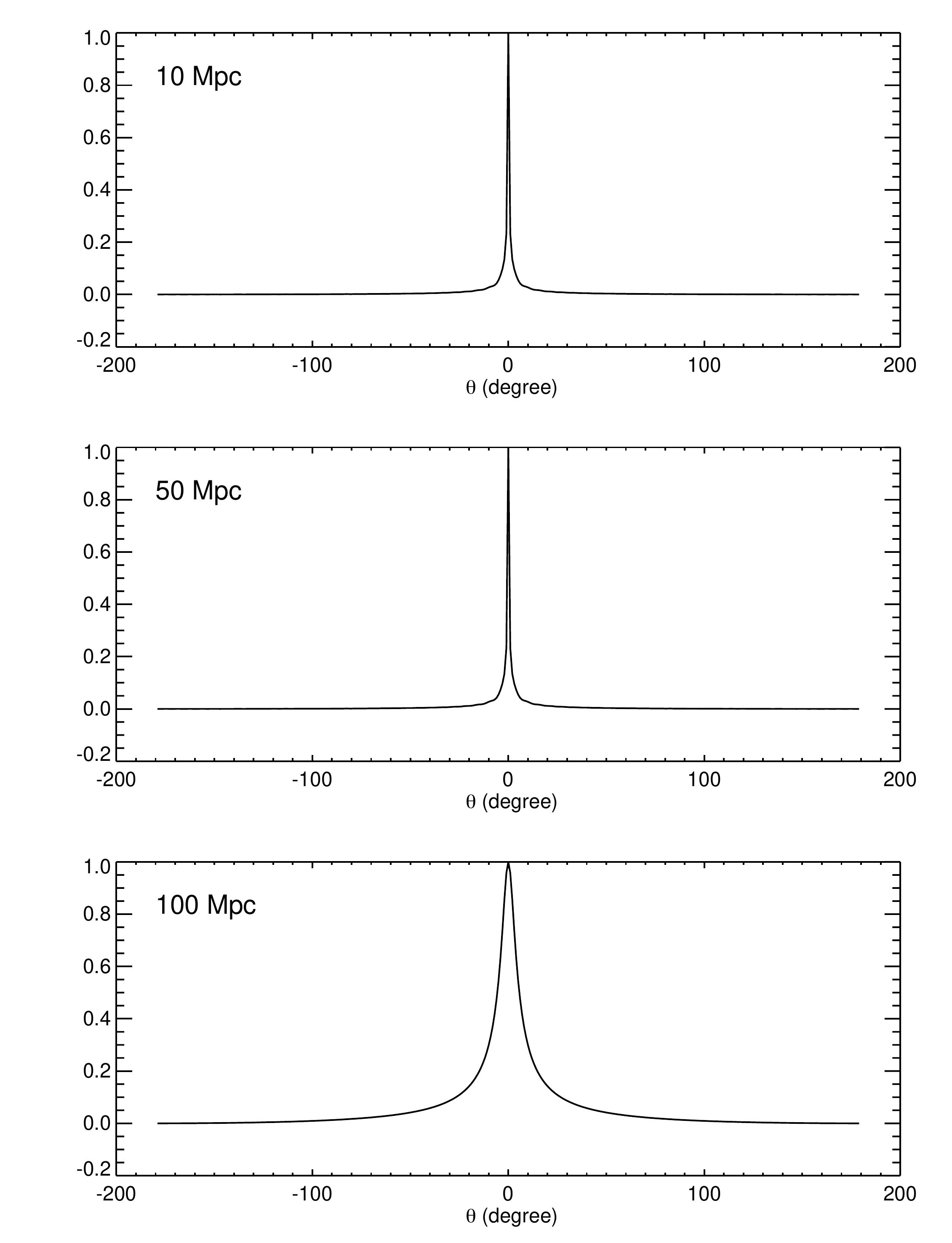}
        \includegraphics[width=0.3\linewidth]{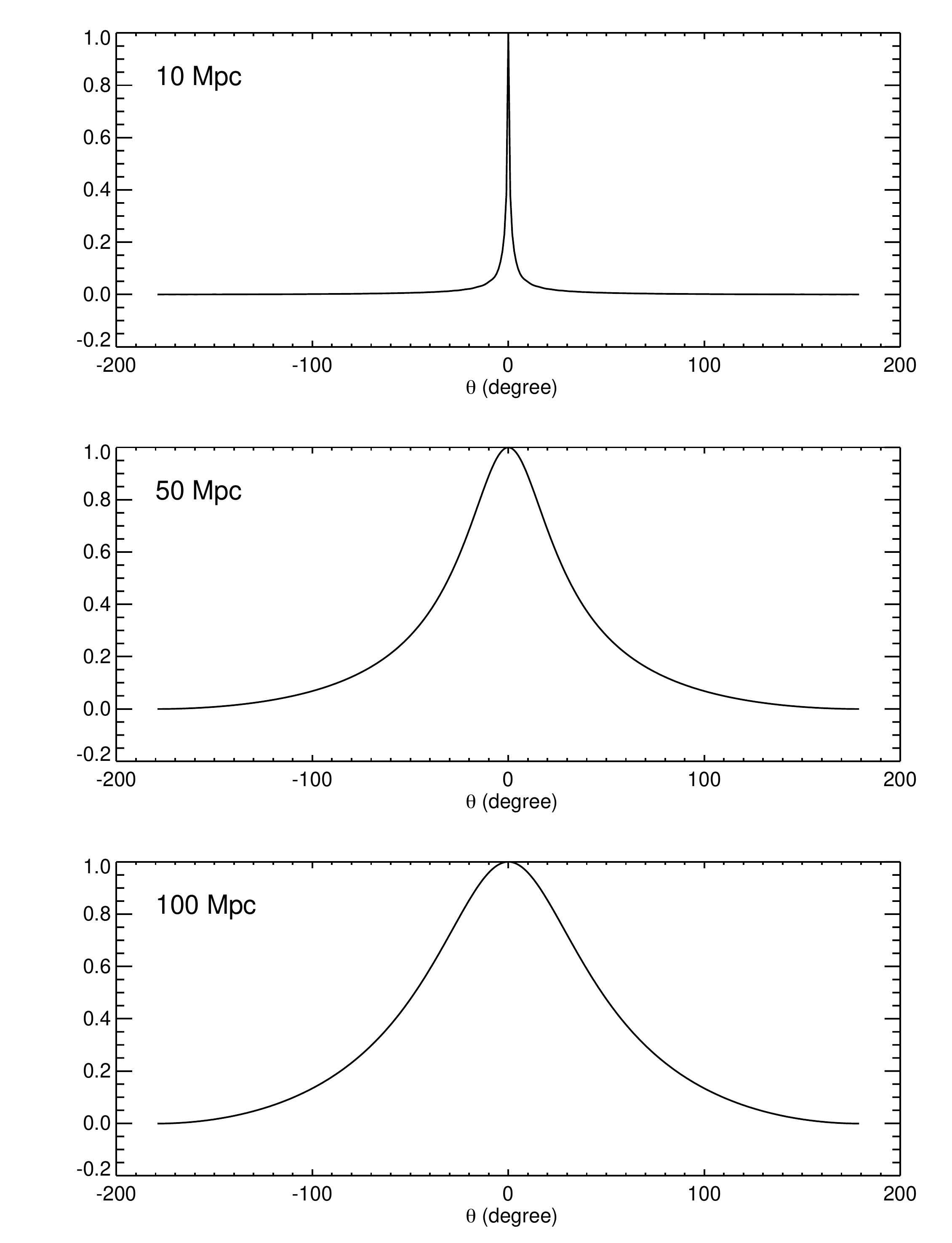}
\caption{The PSF (Eq.~\eqref{eq:psf}) of a single source as a function of distance, for (from left to right) 0.3nG, 0.5nG, and 1 nG, for nitrogen ($Z=7$) at 11.5 EeV (this is the median value of the energy bin $\geq 8$ EeV), i.e. a rigidity $\sim$1.6 EV. One can see that for our best fit parameter model ($\sim$0.3 nG) the spreading due to the EGMF is negligible.}
\label{fig:PSF}
\end{figure*}

A general phase function that holds for inhomogeneities of various sizes is the \citet{Henyey:1941gpa} phase function:
\begin{equation}
P(\cos\delta\theta)=\frac{1-q^2}{\left(1-2q\cos\delta\theta+q^2\right)^{3/2}}
\label{eq:phase_function}
\end{equation}
where  $q=\braket{\cos\delta\theta}$. In the context of cosmic-ray scattering on magnetic field inhomogeneities,  $\delta\theta\sim(\lambda/r_L)^\kappa$ (see above) and therefore we have two cases:
\begin{itemize}
    \item in the weak scattering regime, $\delta\theta<<1$ and $q\sim1$ (small angle scattering); 
     \item in the resonant scattering regime, $\delta\theta\sim1$ and $q\sim0$ (large angle scattering).
\end{itemize}
The intensity of a single source in the sky can be calculated using Eqs.~\eqref{eq:single_source_intensity}-\eqref{eq:phase_function}: 
\begin{equation}
I(\tau,\mu)=\Sigma_{m=0}^{\infty}\left( g_m(\tau)+g_{m+1}(\tau) \right)L_m(\mu)
\label{eq:psf}
\end{equation}
where $I(\tau,\mu)$ is the intensity of an isotropic point source for a given radial direction $\theta$,
$L_m(\mu)$ is the Legendre polynomial of order $m$, $g_m(\tau)=I_0\exp(-\beta_m\tau-\alpha_m\log\tau)$,
$\beta_m=(2m+1)/m(1-q^{m-1})$, $\alpha_m=m+1$ and $g_0(\tau)=0$ \citep[see][for more details]{narasimhan2003shedding}. The number of coefficients used in our calculations is $m=300$. The resulting $I(\tau,\mu)$ corresponds to the point spread function (PSF) that describes the response of the EGMF to a point source (see eamples in Fig.~\ref{fig:PSF}). We use this PSF to calculate the observed intensity map for each shell of distance after diffusion in the EGMF  $I_R(r,\hat{e})$:
\begin{equation}
I_R(r,\hat{e})=I_0(r,\hat{e}')*{\rm PSF}(r,D_{\rm EG},\delta\theta),
\label{eq:shell_image}
\end{equation}
 where $*$ is the convolution operator and  $I_0(r,\hat{e}')$ is the original "shell image" at a distance $r$. 
 
Finally, the "illumination map", $I_j(\hat{e},D_{\rm EG})(E)$, the map of the total UHECR flux illuminating the Galaxy for a composition $j$ at energy $E$ is obtained by summing the intensity sky map for each shell (Eq.~\eqref{eq:shell_image}), with the weighting determined by energy losses and composition evolution. 

\subsection{UHECR horizon} \label{Appendix:horizon}
In this work we have employed two methods of weighting the shells.  The first method is that used by \citet{GPH18}, in which all shells up to a maximum distance (which they designated the UHECR horizon, $H_j(E)$) are given equal weight, and shells at larger distance make zero contribution.  An improvement on this ``sharp cutoff" treatment, short of a full analysis, is the ``exponential attenuation'' of Eq.\eqref{eq:d90} discussed in Sec.~\ref{subsec:source_horizon}.\\

\noindent \underline{\it ``Sharp cutoff'' treatment:} \\ The following is the treatment of \citet{GPH18}: 
With no EGMF the UHECR horizon is taken to be the mean total energy loss length, 
$\chi_{\rm loss}=-c(d\ln E/dt)^{-1}$, 
including the contribution of pair production and photodisintegration processes \citep[e.g.][for more details about the calculations]{Allard:2006mv}. The energy loss length is calculated using the giant dipole resonance cross-sections discussed in \citet{Khan:2004nd} and for the higher-energy processes (pion production), the parameterization of \citet{Rachen1996}.

\begin{figure*}
\centering
    \centering
         \centering
         \includegraphics[width=0.5\linewidth]{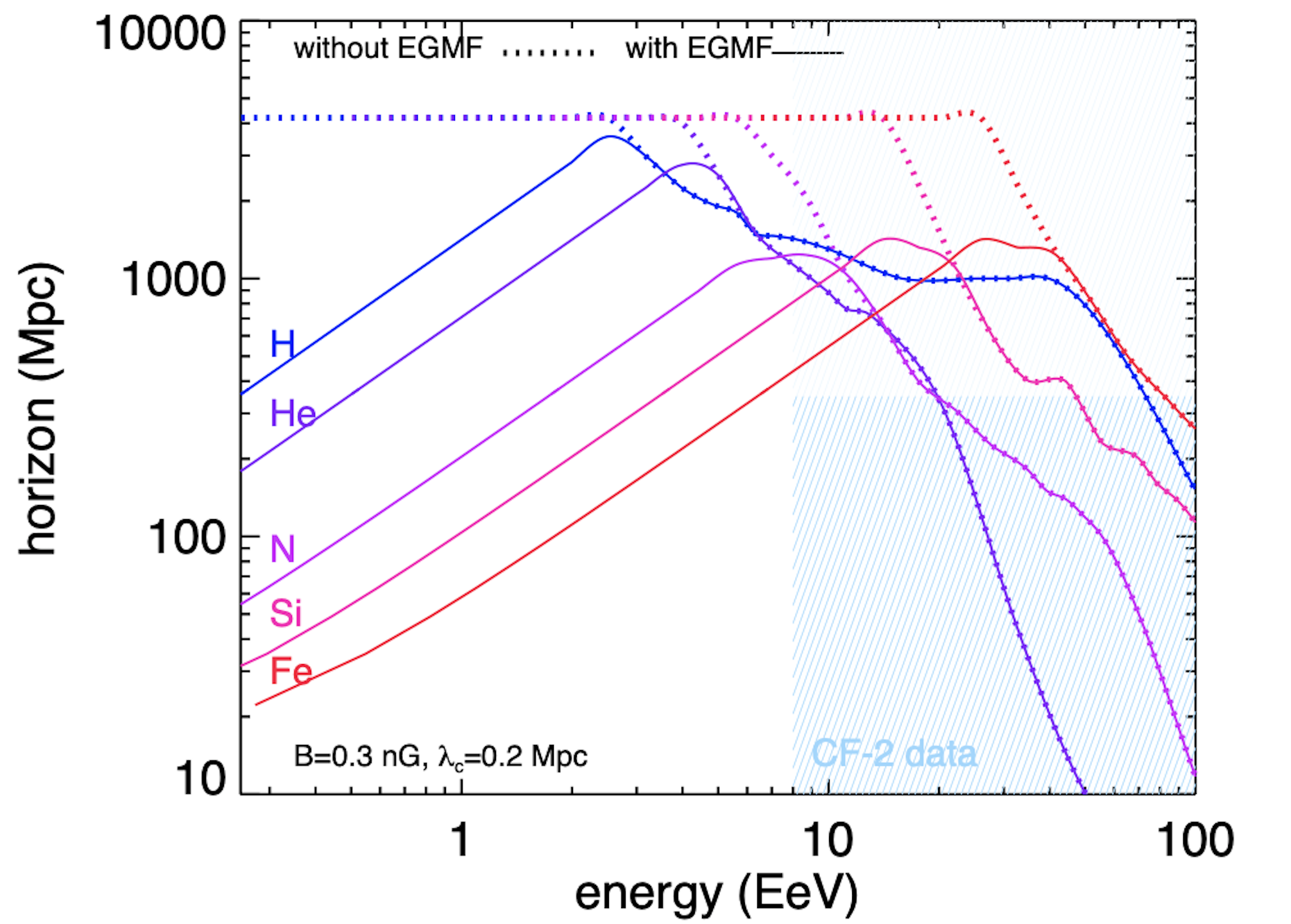}
\caption{UHECR horizon (energy loss length) as a function of energy, for the best fit parameters of the SH* case, $B_{EG}\sim0.3$ nG and $\lambda_{EG}=0.2$ Mpc, for 5 different nuclei $j$: H, He, N, Si, Fe. The extent of the \textit{CosmicFlow-2} reconstruction of the local universe is figured by the blue shaded area. For comparison, the case of zero EGMF is also shown. }
\label{fig:horizon_flux}
\end{figure*}

For a nucleus $j$ with energy $E$ denote
\begin{equation} 
H_j(E) \approx \min(\sqrt{ d_{\rm diff} d_{\rm GZK}} ,d_{\rm GZK}) \ , 
\label{eq:H}
\end{equation} 
where  $d_{\rm GZK}$ is taken to be the energy attenuation length, $\chi_{\rm loss}$, the distance within which a cosmic ray loses $\sim$63\% of its initial energy \citep{GPH18}, and 
 $d_{\rm diff} \sim 6 D_{\rm EG}/c$ \citep{GAP08}
 is the rigidity-dependent diffusion distance. 
For the rigidity range considered, the propagation of cosmic rays is almost ballistic  (i.e. the diffusion coefficient is given by the right hand side of Eq.~\eqref{eq:dcoef}) so the pathlength of the UHECR is approximately the rectilinear distance to the source \citep[see also ][for more details]{GAP08}. 
Fig.~\ref{fig:horizon_flux} shows the horizon $H_j(E)$ from Eq.~\eqref{eq:H} for the best fit parameters.  
We calculate illumination sky maps $I_j(\hat{e},D_{\rm EG})(E)$, for different composition $j$ ($Z=1,2,6,8,14,26$) and diffusion coefficient $D_{\rm EG}$, in 25 rigidity bins.\\



\begin{figure}
\centering
\includegraphics[width=0.4\linewidth]{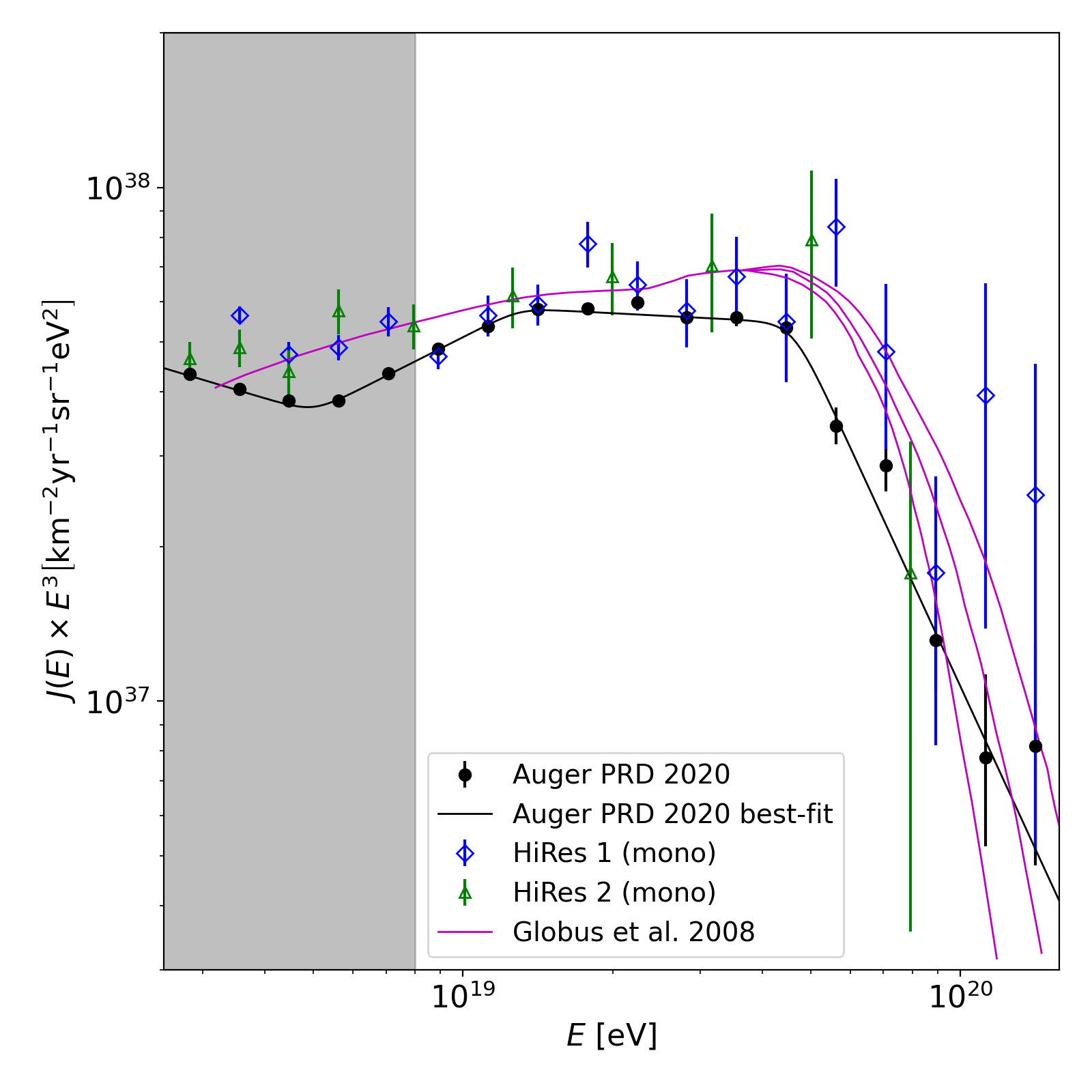}
\caption{The most up-to-date UHECR spectrum from the Pierre Auger Observatory \citep{Auger_PRL_2020, Auger_PRD_2020} (black dots) with the Auger multi-power-law best fit shown by the black line. We take the Auger fit of the energy spectrum as an input to the model. The blue and green markers show the measurement by High Resolution Fly's Eye (HiRes) \citep{HiRes}, with the pink lines being the model of \citet{GAP08} which was guided by the HiRes data and was the source of our $d_{90}$($A,E$) values, showing generally good agreement apart from a small offset that is consistent with energy-scale uncertainties.  The region of the plot below 8 EeV is shaded because it is not relevant to this paper.}
\label{fig:GAP08_spectrum}
\end{figure}

\noindent \underline{\it ``Exponential attenuation'' treatment:}\\  Our second approach to attenuation, the ``exponential attenuation'' treatment, weights the shells by Eq.~\eqref{eq:d90}, where $d_{90}(A,E)$ is the distance within which 90\% of the parents of an observed CR of $(A,\,E)$ originated. 
We take $d_{90}(A,E)$ from \citet{GAP08} which adopted a spectral index of -2.4 and mixed composition at the source, as in \citet{Allard:2005ha}. 
Fig.\ref{fig:GAP08_spectrum} shows that the 2005 HiRes spectrum used by \citet{GAP08} is similar to the most up-to-date Auger spectrum \citep{Auger_PRL_2020, Auger_PRD_2020} imposed in our analysis.

The Auger collaboration \cite{Auger_combined_fit} performed a combined fit to spectrum and composition and found that the best-fitting spectral index depends on the HIM used to interpret the $X_{\rm max}$ data, as well as on the extragalactic background light and photo-interaction models used.  With EPOS-LHC they find a production spectral index $\approx -1$ while other models are compatible with a softer index around $-2$. It would be desirable to test the sensitivity of our results to the d90 modeling, by using  $d_{90}(A,E)$ tables for the combined fit to  spectrum and composition at the source performed by \citet{Auger_combined_fit}, but these are not available in the literature.   We remark that a bias from using integrated rather than binned $d_{90}(A,E)$ tends to offset that of using a too-soft production spectral index in generating the $d_{90}(A,E)$ tables:  If the spectrum is actually harder than -2.4, the true $d_{90}(A,E)$ values will be smaller than adopted, while 
$d_{90}(A,E)$ for events integrated over the spectrum above $E$ is smaller than $d_{90}(A, {\rm lg}E)$ for the single lg$E$ energy bin at $E$, because the horizon decreases with energy.  Thus our procedure using the \citep{GAP08} $d_{90}$ tables, which may use a too-soft spectral index, should be relatively insensitive to the spectral index assumption.  Since the $d_{90}$ approach is inherently inadequate in any case, we have chosen not to invest energy in producing alternate tables ourselves as would be needed to explore the sensitivity to the assumed source spectrum and composition embedded in the $d_{90}(A,E)$'s used.  A complete, correct treatment must self-consistently model the spectrum and composition at the source and treat attenuation of nuclei as motion in $\{A,\,E\}$ space during propagation.

\section{Arrival map calculation procedure}
\subsection{Arrival map calculation procedure} \label{Appendix:arrival}
Once the illumination map is known, one can build the arrival map using the calculation of cosmic-ray trajectories in the GMF model of \cite{JF12}. \citet{FS} provides nearly two billion simulated backtracking trajectories for GMF coherence lengths $\lambda_{\rm coh} = 30$ and 100 pc, with RK4 precision of $10^{-8}$. In order to have more choices of $\lambda_{\mathrm{G}}$, we mix the trajectories of 30 parsec and 100 parsec with different weights to interpolate between them.  

We pixelate the sky map with HEALPix pixels, as they are equal-area pixelation of the sphere. We use HEALPix resolution 5, such that there are 12288 pixels in the sky. The mean spacing of pixels is $1.8323^\circ$, which is fine enough for modelling purpose. Each `source-arrival-matrix' $M_{hi}^{kn}$ (described below) is about 1.2 GB in file size, which is suitable for the memory space allowed by high-speed computation.

Below is the exact procedure of calculating the model arrival map from the illumination map. Start with the following notations:
\begin{itemize}
    \item Model parameter labels $\{m,n,q\}$ label which value of $D_{\rm EG}$, $\lambda_c$ and composition parameter $\boldsymbol{\Omega}$ is discussed.
    \item $S_{h}$ denotes the illumination map, i.e. the cosmic ray flux illuminating the surface of the Milky Way in HEALPix direction $h$. 
    \item $A_{i}$ denotes the arrival map, i.e. the cosmic-ray flux arriving at Earth in HEALPix direction $i$. 
    \item $M_{hi}^{kn}$ denotes the `source-arrival-matrix' that contains information about deflection of cosmic rays in the GMF. The matrix elements are the the numbers of trajectories from source direction $h$ to arrival direction $i$ for rigidity $k$ and GMF coherence length $n$. 
    \item $V_{jk}^{lq}$ denotes the rigidity spectrum, i.e. the abundance of cosmic rays $\frac{\mathrm{d}\,N}{\mathrm{d}\,\lg R}$ of chemical element $j$, rigidity bin $k$, energy threshold $l$ for composition parameter $q$. It is derived from the energy spectrum and the composition parameter as discussed in Sec.~\ref{Appendix:composition}.
    \item $m$ denotes the EGMF diffusion coefficient.
    \item $\omega_i$ denotes the directional exposure of the Pierre Auger Observatory in HEALPix direction $i$.
\end{itemize}

The arrival map of each cosmic ray species in each rigidity bin can be calculated from the illumination map and `source-arrival-matrix' by
\begin{equation}
    A_{i}^{jkmn} = \sum_{h} S_{h}^{jkm} M_{hi}^{kn}.
\end{equation} 

By combining the individual arrival maps with weights given by the rigidity spectrum, the arrival map of all cosmic rays in energy bin $l$ with parameter set $mnq$ is 
\begin{equation}
   \mathcal{A}_{i}^{lmnq} = \sum_{jk} A_{i}^{jkmn}\frac{V_{jk}^{lq}}{\sum_i \omega_i A_{i}^{jkmn}}.
\end{equation}

The $\omega_i$ is included in the denominator to ensure that one would obtain an energy spectrum exactly as observed by Auger with partial exposure, in case the full-sky spectrum is slightly different to the spectrum observed by Auger with partial exposure.

Finally, the illumination map of all cosmic rays is:
\begin{equation}
   \mathcal{S}_{h}^{lmnq} = \sum_{jk} S_{h}^{jkmn}\frac{V_{jk}^{lq}}{\sum_i \omega_i A_{i}^{jkmn}}.
\end{equation}
\subsection{Reconstructing the dipole component from sky map with partial exposure}
There are two prominent ways to reconstruct a dipole from an observed data set with partial exposure: the Rayleigh analysis and the K-inverse method. We tested both in earlier works \citep{PoS19}. The two methods give similar results. The differences between Rayleigh analysis and K-inverse method are mainly manifested in $d_z$. This is expected because the directional exposure is zero for declination above 45 degree and it is difficult to measure $d_z$ with such incomplete coverage. For the Auger data set, the Rayleigh analysis is preferred because it is less sensitive to the systematics of the experiment. However, since we are working from a full-sky model it is most straightforward to use the K-inverse method to reconstruct the dipole which 
would be inferred using the Rayleigh method, given Auger's limited exposure. Table~\ref{tab:result_above8} gives the dipole components of both the full-sky model arrival map and the dipole components reconstructed from the model arrival map weighted with Auger's exposure.

\subsection{Top hat smoothing procedure} \label{sec:top_hat_method}

To smooth the flux map with a 45 deg top hat function for plotting, a standard procedure, used in \citet{Auger_4bin_2018}, consists in dividing the number of events in a region (45 deg radius disk excluding the zero-exposure region) by the total exposure in that region. An alternative way consists in weighting each event by the inverse exposure for its arrival direction, summing the weight of events in the region and dividing by the solid angle of the region (45 deg disk excluding zero-exposure region) as in \citet{Deligny:2015vol}. We use this alternative way, which leads to some differences. In particular, the excess in the region $180^\circ < l < 210^\circ$ and $30^\circ < b < 60^\circ$ appearing in the Auger top-hat map $\geq 32$ EeV \citep{Auger_4bin_2018} may be an artifact as it does not appear in the sky map created using the \citet{Deligny:2015vol} smoothing method applied to our backed-out events above 38 EeV, nor is such an excess present in the LSS model map in Fig.~\ref{fig:dip_AM}.


\section{The effect and uncertainty of the Jansson-Farrar-2012 GMF model} \label{Appendix:JF12}

In this section we discuss the JF12 model uncertainties. UHECRs from a given direction receive a net deflection from the ordered component of the field as well as being spread out by the turbulent component. Besides deflecting and diffusing the arrival directions of cosmic rays, the GMF can amplify or reduce the flux at Earth, depending on the direction of cosmic rays entering the Galaxy and their rigidity.  This effect is naturally incorporated in the \citet{FS} trajectories.

Being more general and constrained by more data, JF12 is the generally adopted model of the coherent GMF. A strong point of JF12 relative to other GMF models is that JF12 used polarized synchrotron emission and not just RMs for constraining the model parameters, thus constraining the transverse as well as line-of-sight components of the magnetic field.  However interpreting the all-sky RM, Q and U measurements requires 3D models of the thermal and cosmic ray electron distributions, and the models available in 2012 (and still now) are rather primitive, limiting the accuracy of the inferred coherent GMF.  The JF12 random field model was based on unpolarized synchroton emission maps from WMAP.  Since then, Planck has found the unpolarized synchroton emission to be considerably less than WMAP reported, \citet{planckB}
due to a different treatment of other contributions.  This could potentially lead to a substantial reduction in the magnitude of the random component of the GMF. (For further discussion see \citet{fCRAS14} and \citet{planckB}.) While we do not have trajectories simulated in a weaker random field, functionally speaking the $\lambda_c = 30$ pc simulations of \cite{FS} have a similar qualitative effect to a reduced field strength.   

\section{Fitting methodology}\label{Appendix:fitting}

\subsection{Composition modelling and likelihood} \label{Appendix:composition}

The mass composition of UHECRs can be inferred from measurements of the maximum atmospheric depth of air showers, $X_{\max}$, by the Auger collaboration \citep{ICRC2019_Auger_comp}. The mean and variance of $\ln A$ ($A$ is the mass number) and their uncertainties at 8 energy levels, are inferred using three different HIMs: Sibyll2.3c, EPOS-LHC, QGSJETII-04. 

Even though we are informed by $X_{\max}$ measurements, given the uncertainty on the composition from the HIMs we parameterize the evolution of the composition with energy and take the $X_{\max}$ derived composition as constraints along with the dipole anisotropy. 
We parameterize the composition as follows (abbreviating ${\rm log}_{10}(E) \equiv \lg E$):
\begin{itemize} \vspace{-0.07in}
    \item $\langle\ln A\rangle=\alpha \lg E + \langle\ln A\rangle_{8\rm\, EeV}$ \vspace{-0.07in}
    \item $\sigma^2(\ln A)=\beta \lg E +{\sigma^2(\ln A)}_{8\rm\, EeV}$. \vspace{-0.07in}
\end{itemize}
Thus the model composition is characterized by four parameters $\boldsymbol{\Omega}\equiv  \{\alpha, \beta,  \langle\ln A\rangle_{8\rm\, EeV}, {\sigma^2(\ln A)}_{8\rm\, EeV}\}$. 

The following discussion is about how to calculate $\ln L(\langle\ln A\rangle\mid\boldsymbol{\Omega};\mathrm{HIM}) + \ln L(\sigma^2(\ln A)\mid\boldsymbol{\Omega};\mathrm{HIM})$, i.e. the log of likelihood that the true mean and variance of $\ln A$ at 8 energy levels are indicated by $\boldsymbol{\Omega}$, given that a particular HIM is true.

Let us use the following notations:
\begin{itemize}
    \item $N$ is the total number of energy bins of $X_{\max}$ measurement relevant to this study, in this case $N=8$. The $\langle\lg E\rangle$ are 18.95, 19.05, 19.15, 19.25, 19.35, 19.44, 19.55, 19.73.
    \item $\boldsymbol{\Omega_1} \equiv \{\alpha, \langle\ln A\rangle_{8\rm\, EeV}\}$ the set of two parameters regarding $\langle\ln A\rangle$.
    \item $\mathbf{x}$ is the $\langle\ln A\rangle$ of cosmic ray in $N$ energy bins inferred by a chosen HIM.
    \item $\mathbf{m}$ is the $\langle\ln A\rangle$ of cosmic ray in $N$ energy bins specified by $\boldsymbol{\Omega_1}$.
    \item $\mathbf{s}$ is the systematic uncertainty of $\mathbf{x}$ in $N$ energy bins of the chosen HIM.
    \item $\boldsymbol{\sigma}$ is the statistical uncertainty of $\mathbf{x}$ in $N$ energy bins of the chosen HIM.
    \item $\mathbf{C}$ is the covariance matrix $C_{ij}= \sigma_{i}^{2} \delta_{i j}+s_{i} s_{j}$. 
\end{itemize}

For each HIM and each parameter set $\boldsymbol{\Omega_1}$,
\begin{subequations} \label{eq:covariance_approach}   
\begin{align}
    L(\langle\ln A\rangle\mid\boldsymbol{\Omega};\mathrm{HIM}) &\equiv p(\mathbf{x}\mid\mathbf{m}) = B \exp(-\chi^{2}_{\min}/2) \\
    \chi^{2}_{\min} &= (\mathbf{x}-\mathbf{m})^{\mathrm{T}}\mathbf{C}^{-1}(\mathbf{x}-\mathbf{m}) \\
    B &= {\left[(2 \pi)^{N}\operatorname{det} \mathbf{C}\right]}^{-1/2} \\
    \ln L &=-\frac{1}{2}\left[(\mathbf{x}-\mathbf{m})^{\mathrm{T}} \mathbf{C}^{-1}(\mathbf{x}-\mathbf{m})+\ln (\operatorname{det} \mathbf{C})+ N \ln (2 \pi)\right] 
\end{align}
\end{subequations}
The lower and upper systematic uncertainty of $\langle\ln A\rangle$ are not equal, such that $\mathbf{s}$ and $\mathbf{C}$ are not uniquely defined. So we introduce a nuisance parameter $\eta$ such that the systematic shift of $x_{i}$ is $\eta s_{i}$. With either upper or lower systematic uncertainty, one can calculate 
\begin{equation}
    \chi^{2}_{\min} = \min_{\eta} \left( \sum_{i=1}^{N} \frac{\left(m_{i}-x_{i}-\eta s_{ i}\right)^{2}}{\sigma_{i}^{2}}+\eta^{2} \right) \label{eq:eta}
\end{equation}
and compare the two $\chi^{2}_{\min}$ to determine if the lower or upper systematic uncertainty should be applied. Then $\mathbf{C}$ is uniquely defined and one can proceed to calculate $\ln L$ using Eq.~\eqref{eq:covariance_approach}.

The calculation procedure of $\ln L(\sigma^2(\ln A)\mid\boldsymbol{\Omega};\mathrm{HIM})$ is similar to that of $\ln L(\langle\ln A\rangle\mid\boldsymbol{\Omega};\mathrm{HIM})$. The only difference is that the lower and upper \textit{statistical} uncertainty of $\sigma^2(\ln A)$ are not equal either. The calculation takes more effort, but the concept is clear: first use Eq.~\eqref{eq:eta} to determine if upper/lower \textit{systematic} uncertainty should be used for all $x_i$ and determine if upper/lower \textit{statistical} uncertainty should be used for individual $x_i$, then determine $\mathbf{C}$ and calculate $\ln L$ using Eq.~\eqref{eq:covariance_approach}.

We consider seven chemical elements, p, He, C, O, Ne, Si, Fe, and mix them such that the $\langle\ln A\rangle$ and $\sigma^2(\ln A)$ in each narrow $\lg E$ bin of width 0.02 are as specified by a given choice of $\boldsymbol{\Omega}$. The fractions are in general not unique, but if we use fewer composition bins unphysical discretization effects would be introduced.  To choose among the different solutions, we  
adopt the one whose element fractions have the median skewness of $\ln A$. In the case that certain $\sigma^2(\ln A)$ cannot be achieved, we take the element fractions that give the closest value. We tested this procedure using the all-A composition predictions of the \citet{MUF19} model.  In order to compare with the Auger data of $\langle\ln A\rangle$ and $\sigma^2(\ln A)$ that use wider energy bins, we group narrow energy bins together accordingly.

\subsection{Likelihood of dipole components}
The Auger measurement of dipole components $d_x, d_y, d_z$ of three energy bins 8-16, 16-32 and $\geq 32$ EeV are independent of one another \citep{Auger_4bin_2018}. The formula below is used to calculate the log-likelihood for Auger to measure the dipole component $d_i$ with value $x_i$ and Gaussian uncertainty $\sigma_i$ if the true dipole component $m_i$ is given by the model.
\begin{equation}
    \ln L(\mathrm{dipole}\mid\boldsymbol{\Theta};\mathrm{source}) = \sum_{i=1}^9 \ln \left[\frac{1}{\sqrt{2 \pi} \sigma_i} \exp\left(- \frac{{(x_i - m_i)}^2}{2 \sigma_i}\right)\right]
\end{equation}

\subsection{Likelihood of arrival directions of events above 38 EeV}

To check the compatibility of our LSS model with observations, we performed a hypothesis test using the same test statistic as in \citet{Auger_starburst_gAGN} and \citet{TA_test_starburst} based on the log-likelihood ratio. With an isotropic flux model $\Phi_{\rm ISO}(\hat{\boldsymbol{n}})$ and the model flux model $\Phi_{\rm model}(\hat{\boldsymbol{n}})$ as null hypothesis and alternative hypothesis respectively, the test statistic is defined as twice the log-likelihood ratio

\begin{equation}
\begin{aligned}
\mathrm{TS} &=2 \ln \left(L\left(\Phi_{\rm model}\right) / L\left(\Phi_{\rm ISO}\right)\right) \\
\text{ where } L\left(\Phi\right) &=\prod_{i} \frac{\Phi\left(\hat{\boldsymbol{n}}_{i}\right) \omega\left(\hat{\boldsymbol{n}}_{i}\right)}{\int_{4 \pi} \Phi(\hat{\boldsymbol{n}}) \omega(\hat{\boldsymbol{n}}) \mathrm{d} \Omega},
\end{aligned} 
\end{equation}
and $\hat{n}_{i}$ being the arrival direction of the $i$-th observed event and $\omega(\hat{n})$ being the directional exposure of Auger experiment. We define 
\begin{equation} \label{eq:lnL_event}
\ln L(\mathrm{events}\mid\boldsymbol{\Theta};\mathrm{source}) \equiv \ln \left(L\left(\Phi_{\rm model}\right) / L\left(\Phi_{\rm ISO}\right)\right) = 0.5 \,\mathrm{TS}.
\end{equation}
This way it is the correct log-likelihood that can be added with log-likelihood from other aspects (such as composition) and it is normalized such that $\ln L(\mathrm{events}\mid\boldsymbol{\Theta};\mathrm{source})=0$ if ISO is the model.

\subsection{Reconstructing the arrival directions of events above 38 EeV}

\begin{figure*}
\includegraphics[width=0.49\linewidth]{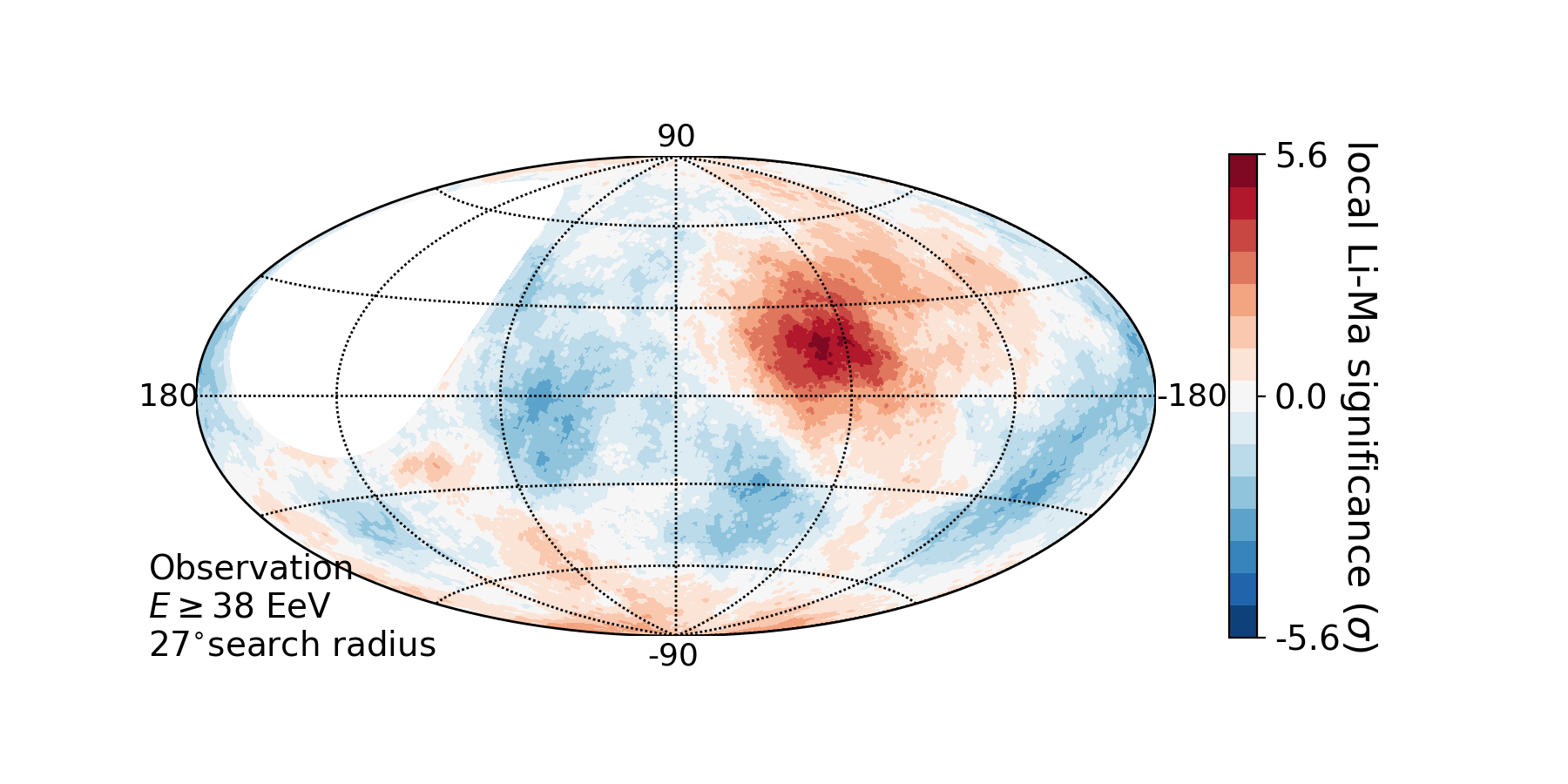}
\includegraphics[width=0.49\linewidth]{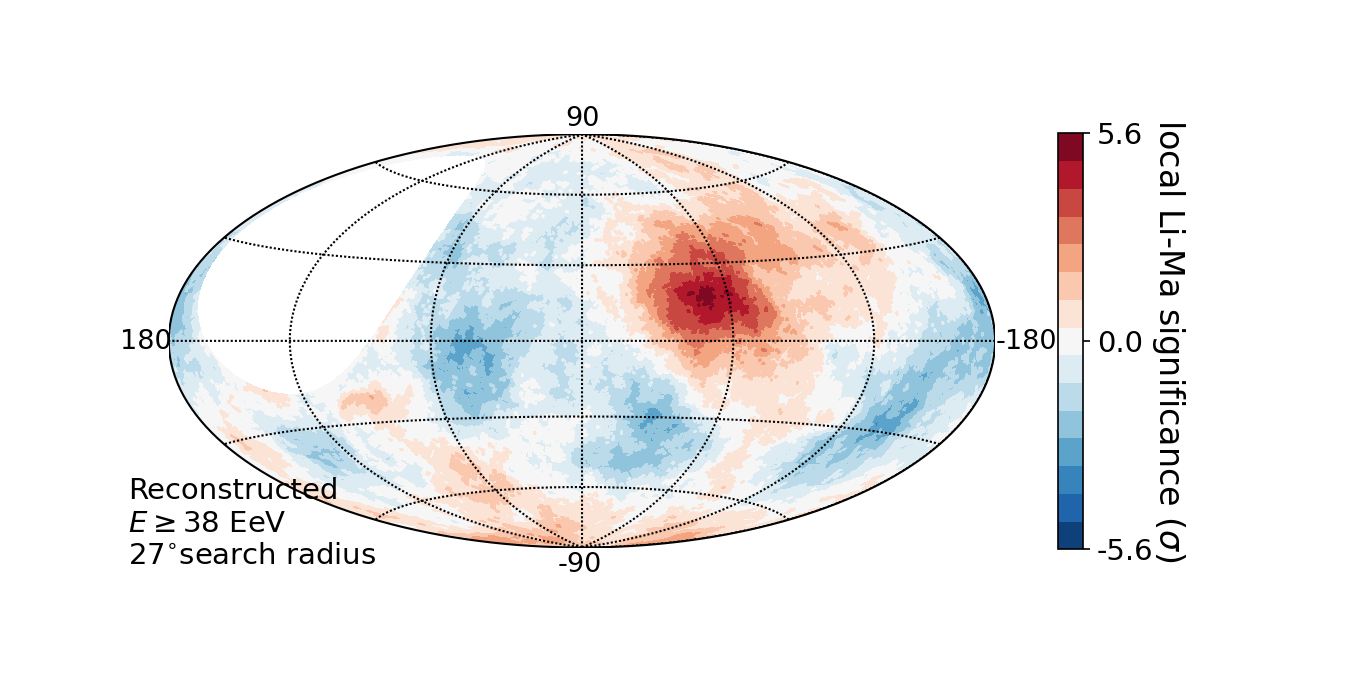}\\
\includegraphics[width=0.49\linewidth]{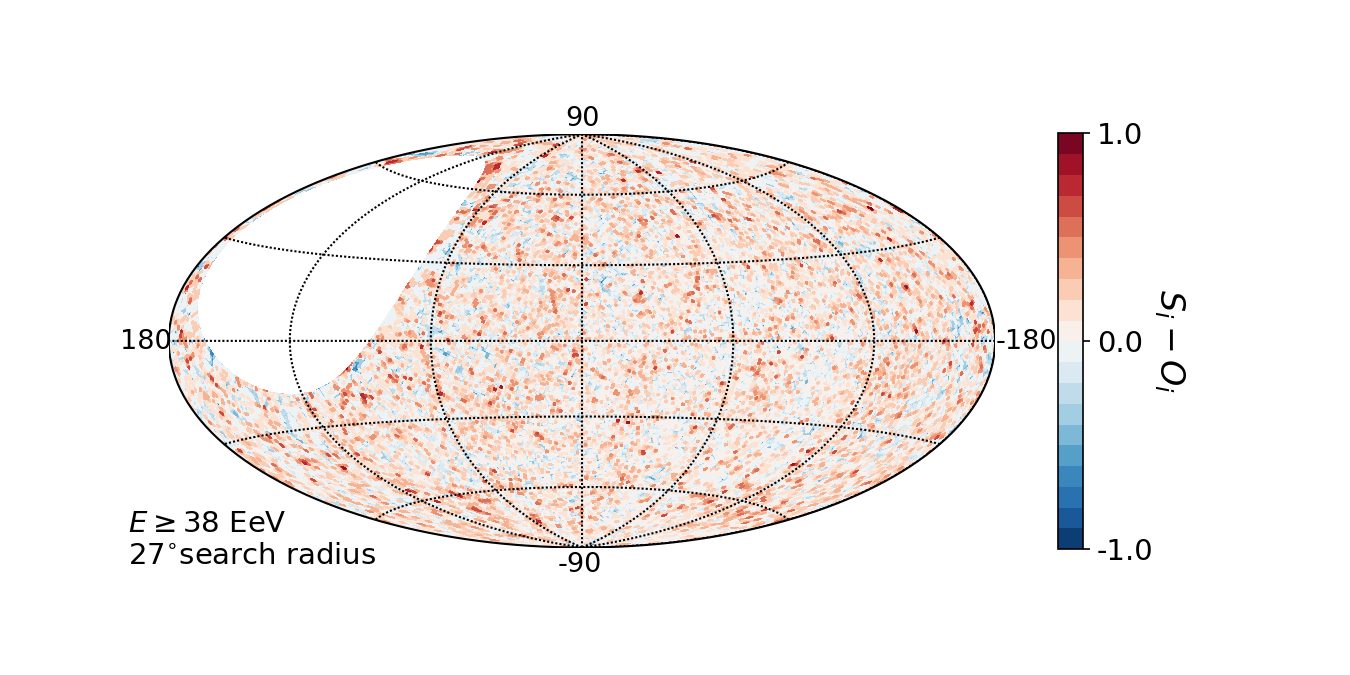}
\includegraphics[width=0.40\linewidth]{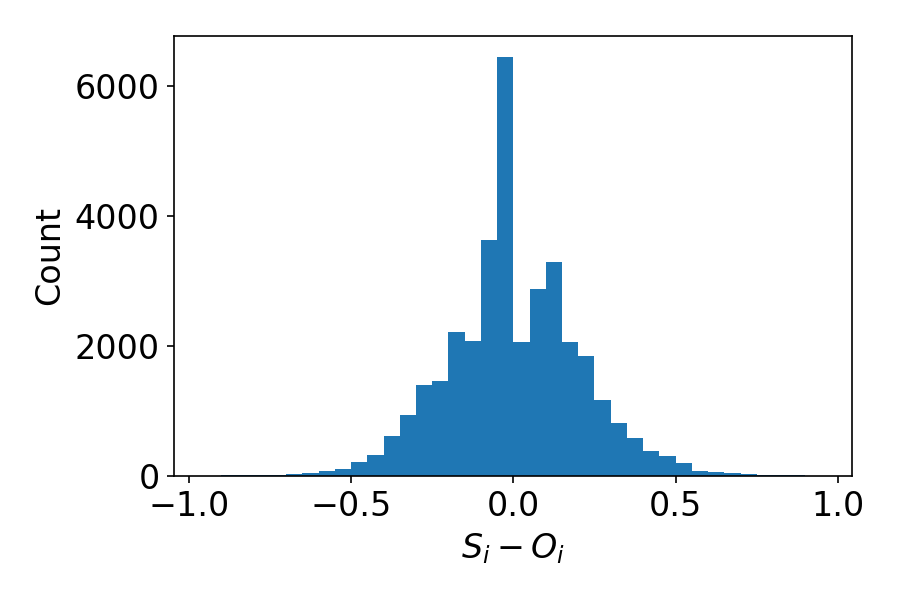}\\
\includegraphics[width=0.49\linewidth]{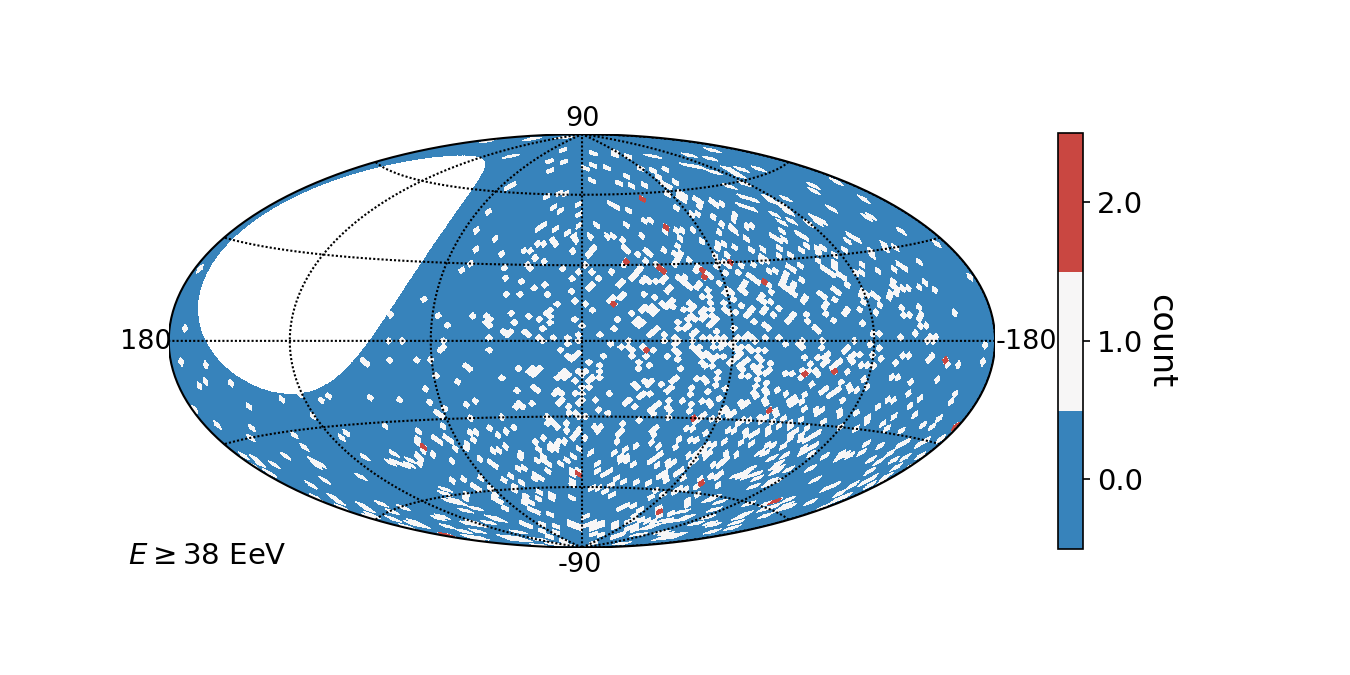} 
\includegraphics[width=0.40\linewidth]{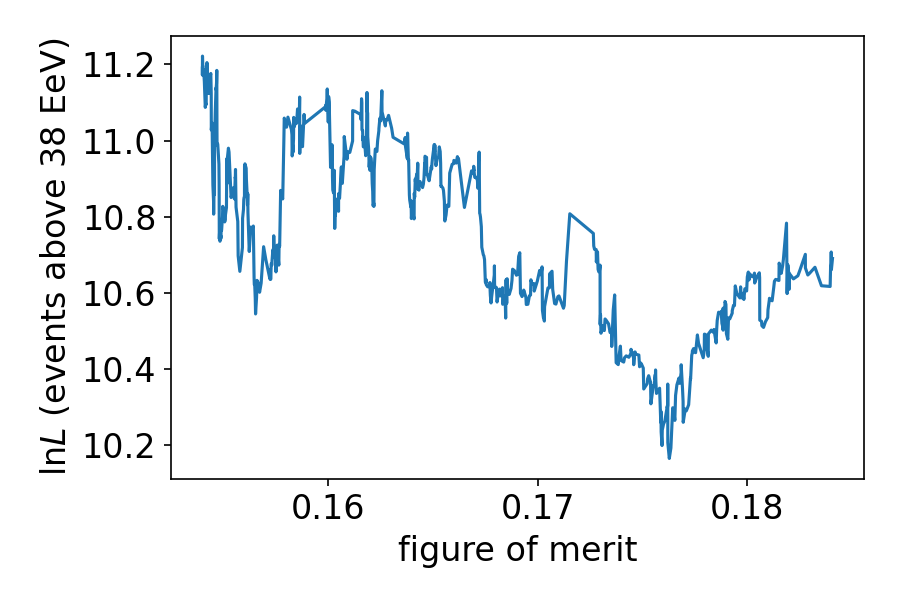}
\caption{Upper plots: the Li-Ma significance sky map from \citet{ICRC2019_Auger_hotspot} (left) and that calculated from reconstructed arrival directions (right). Middle: the sky map (left) and histogram (right) of $S_i-O_i$, i.e. the difference in Li-Ma sky map between observation and reconstruction. Lower left: the sky map of reconstructed arrival directions of the events above 38 EeV. Lower right: the evolution of the value of Eq.~\eqref{eq:lnL_event} for the model case SH* as the figure of merit is minimized. The variation is relatively small compared to its value. A larger $\ln L$ means the model fits better the arrival directions of events above 38 EeV.}
\label{fig:reconstruction}
\end{figure*}

The arrival directions of the events above 38 EeV used for the \citet{ICRC2019_Auger_hotspot} analysis are not yet published, but they are needed because Eq.~\eqref{eq:lnL_event} is the best way to measure the similarity between model and data. We digitized the Li-Ma significance sky map in Fig.~1 of \citet{ICRC2019_Auger_hotspot} and used it to reconstruct the arrival directions of the events. The reconstruction procedure is as follows. We uniformly tiled the sky with 12288 HEALPix pixels of area 3.357 sq deg, so each 27 deg circle consists of a batch of 682 pixels. Let us use the following notations:
\begin{itemize}
    \item $N_{\mathrm{tot}}$ is the total number of events above 38 EeV; $N_{\mathrm{tot}}=1288$.
    \item $n_{\mathrm{pix}}$ is the total number of HEALPix pixels that have non-zero detector exposure; $n_{\mathrm{pix}}=10477$.
    \item $n_{\mathrm{grid}}$ is the total number of data points in the Auger Li-Ma skymap. The data is in 1 degree grid in equatorial coordinates (right ascension and declination) excluding zero exposure region.
    \item $\mathbf{N}$ is a 1D array of length $n_{\mathrm{pix}}$ that represents the number of events in each individual pixel, with the sum of all elements being $N_{\mathrm{tot}}$.
    \item $\mathbf{N}_{\mathrm{on}}$ is a 1D array of length $n_{\mathrm{grid}}$ that represents the number of events inside each 27 degree circle centered at each grid point.
    \item $\boldsymbol{\alpha}$ is a 1D array of length $n_{\mathrm{grid}}$ that represents the probability of an event to fall inside each 27 degree circle if the flux is isotropic; it is effectively the exposure in each circle divided by total exposure.
    \item $\mathbf{A}$ is a 2D matrix comprised of $1$'s and $0$'s such that $A_{ij}$ represents if pixel $j$ belongs to the batch of pixels that represents the circle centered at grid point $i$; the matrix is $n_{\mathrm{grid}}$ by $n_{\mathrm{pix}}$.
    \item $\mathbf{S}$ is a 1D array of length $n_{\mathrm{grid}}$ that represents the Li-Ma significance of a trial data set at each grid point.
    \item $\mathbf{O}$ is a 1D array of length $n_{\mathrm{grid}}$ that represents the Li-Ma significance of observation data at each grid point.
    \item f is the function of Li-Ma significance as in Eqn 17 of \citet{Li-Ma}.
\end{itemize}
The following relationships hold
\begin{subequations} \label{eq:reconstruction}   
\begin{align}
\mathbf{A} \mathbf{N}&=\mathbf{N}_{\mathrm{on}} \\
\mathbf{S} &= f(\mathbf{N}_{\mathrm{on}},N_{\mathrm{tot}},\boldsymbol{\alpha}).
\end{align}
\end{subequations}
We turned the reconstruction of arrival directions into a computation problem. Moving an event from one pixel to another effectively alters $\mathbf{N}$ and subsequently alters $\mathbf{N}_{\mathrm{on}}$ and $\mathbf{S}$. The goal is to find the best $\mathbf{N}$ that minimizes the discrepancy between $\mathbf{S}$ and $\mathbf{O}$. We set the objective function for minimization as
\begin{equation}
\text{figure of merit}=\frac{1}{n_{\mathrm{grid}}}\sum_i^{n_{\mathrm{grid}}}|S_i-O_i|.
\end{equation}

We started with a random set of arrival directions. Then we calculated how much improvement could be made to the figure of merit if any one event was moved from one pixel to another. A movement was accepted if it improved the figure of merit. We kept moving individual events one at a time until no further improvement could be made. The figure of merit reached 0.162 which was a local minimum. In order to further improve the reconstruction, we had to let it get out of the local minimum by moving multiple events at once. We randomly selected six events within an angular distance of 54 degree, each event was allowed to move to any of its four neighboring pixels or stay where it was. We repeated this procedure until the figure of merit reached 0.154. We believed this was sufficiently good for our purpose and we did not spend more computation resources on it. The upper two plots of Fig.\ref{fig:reconstruction} show the Li-Ma significance sky map from \citet{ICRC2019_Auger_hotspot} and that from reconstruction. They look very alike. The middle two plots show the sky map and histogram of $S_i-O_i$. One can see that the excesses and deficits of $S_i-O_i$ are randomly distributed in the sky map. The lower left plot shows the reconstructed arrival directions in a sky map. The lower right plot shows the evolution of the value of Eq.~\eqref{eq:lnL_event} with the best-fit (fixed) parameters of case SH* as the figure of merit is minimized. The variation in the value of Eq.~\eqref{eq:lnL_event} is sufficiently small that it does not affect the results of this paper. Comparing the arrival directions in the best reconstruction with those in previous iterations, the arrival directions of many cosmic rays have not been changed, and those of the rest of the cosmic rays are moved by one or two pixels. The mean spacing of pixels is $1.8323^\circ$, so we estimate that the accuracy of the reconstructed arrival directions is a few degrees. Apart from accuracy, the precision level of the reconstruction is limited by the pixelization of the sky. Having more pixels would increase precision but increase computation cost.


\section{More results details}

\subsection{Contribution from individual shells}\label{Appendix:allshells}

\begin{figure*}
\centering
    \centering
         \centering
         \includegraphics[width=0.33\linewidth,trim=0.5cm 1.5cm 0.5cm 2cm,clip=true]{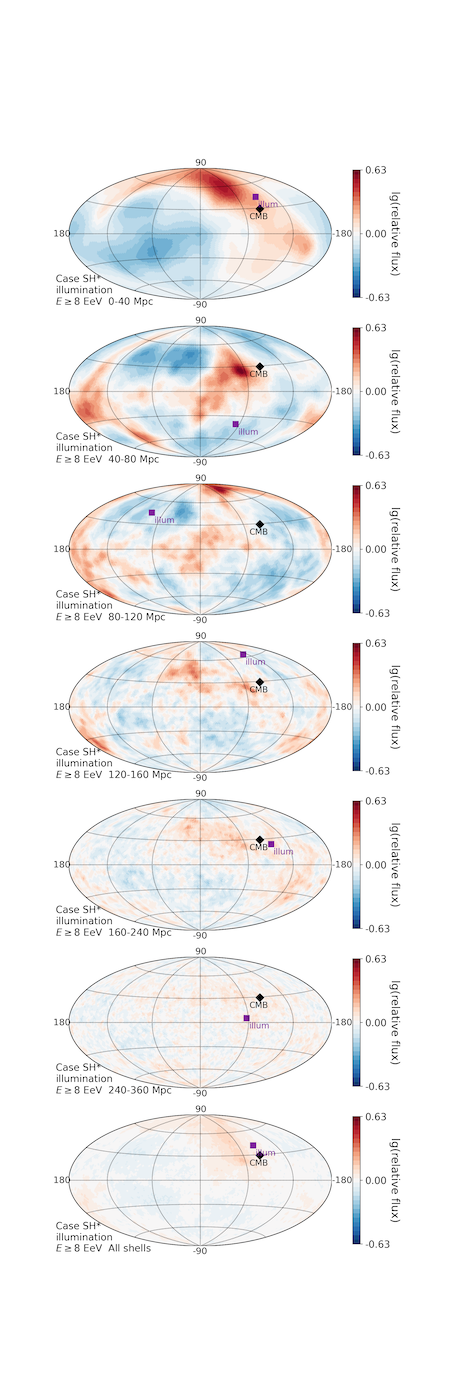}
         \includegraphics[width=0.33\linewidth,trim=0.5cm 1.5cm 0.5cm 2cm,clip=true]{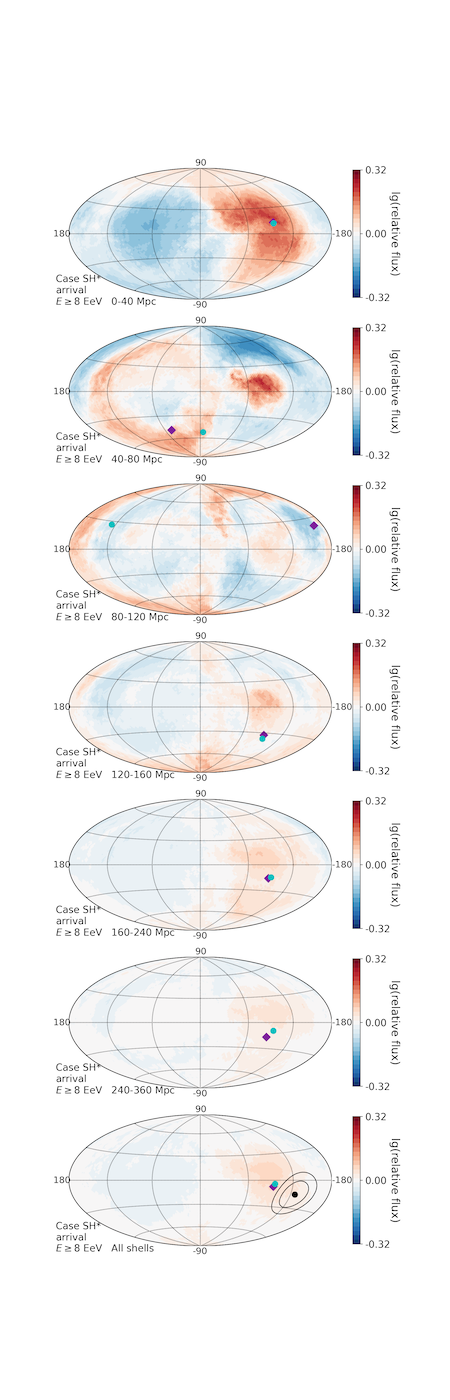}
         \includegraphics[width=0.33\linewidth,trim=0.5cm 1.5cm 0.5cm 2cm,clip=true]{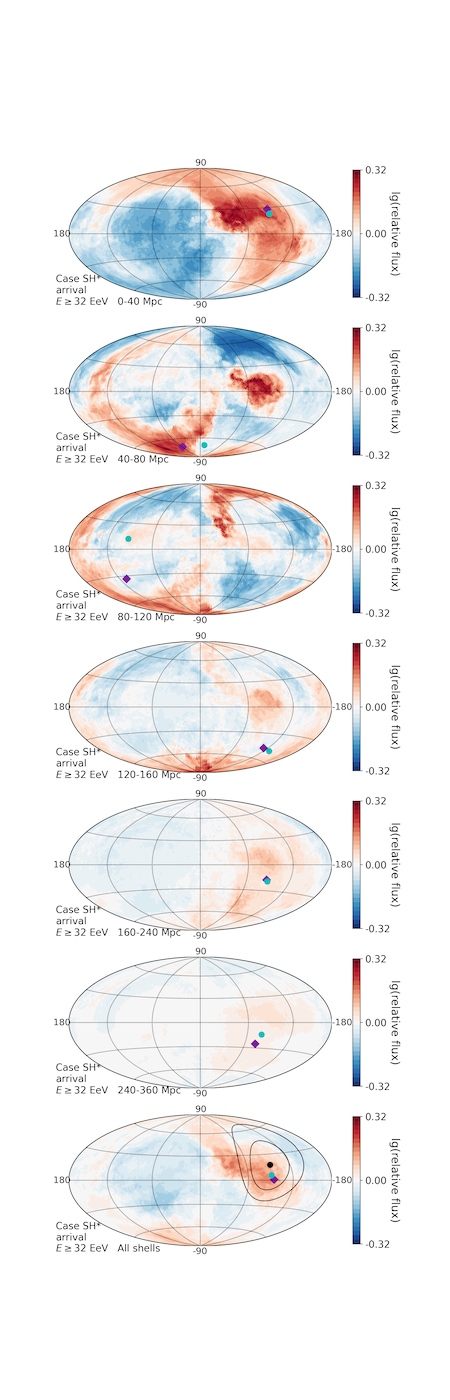}
\caption{Illumination maps (UHECR flux illuminating the Galaxy) and arrival maps (after propagation in the JF12 GMF model) for the model Case ``SH*'' and energy bins $\geq 8$ EeV and $\geq 32$ EeV for different shells of distances.}
\label{fig:allshells_SHasterisk}
\end{figure*}

\begin{figure*}
\centering
    \centering
         \centering
         \includegraphics[width=0.33\linewidth,trim=0.5cm 1.5cm 0.5cm 2cm,clip=true]{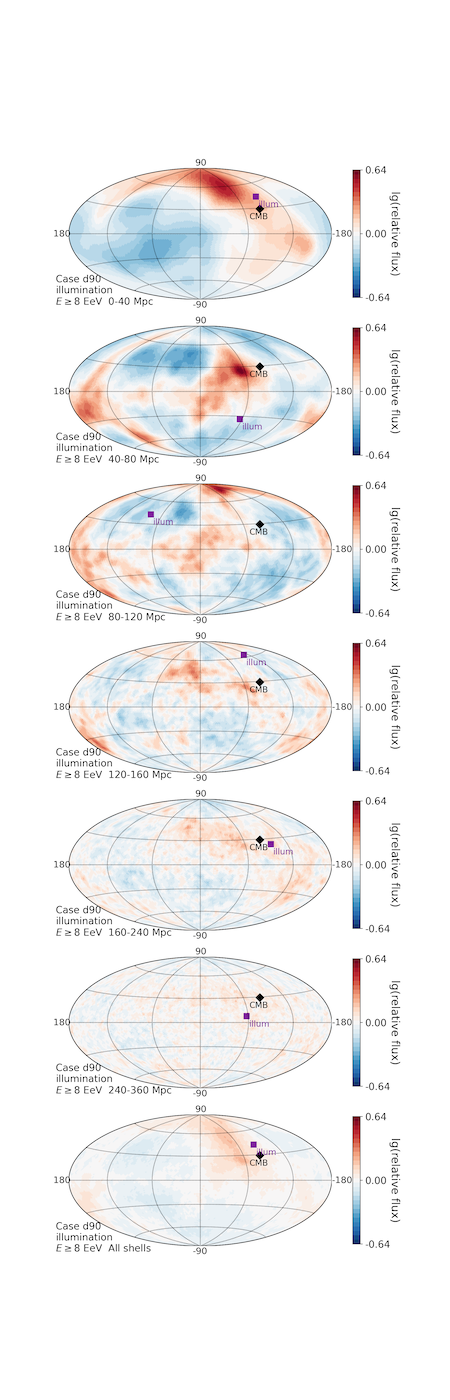}
         \includegraphics[width=0.33\linewidth,trim=0.5cm 1.5cm 0.5cm 2cm,clip=true]{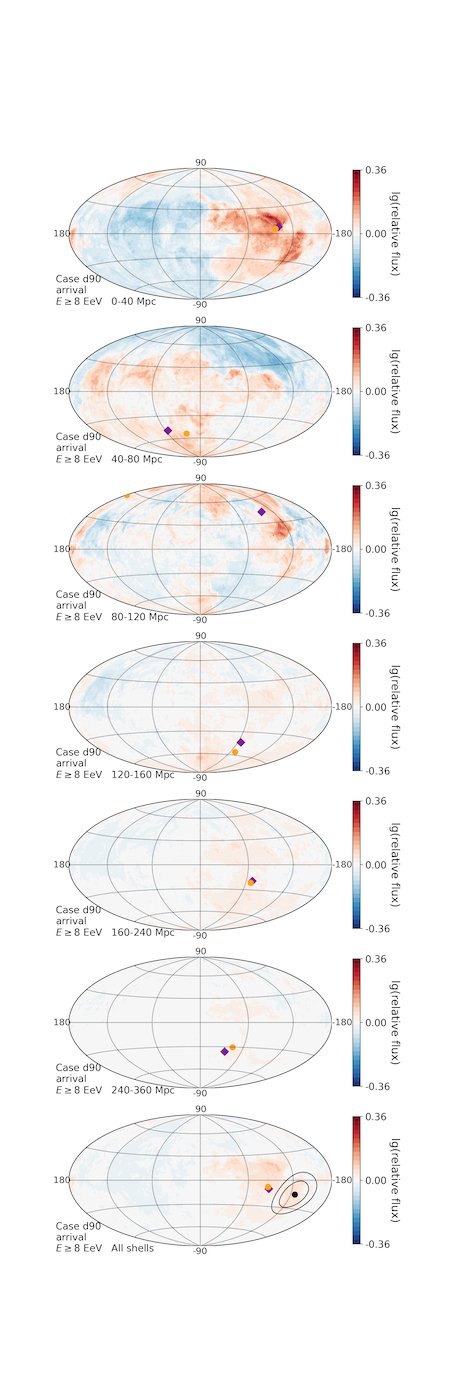}         
         \includegraphics[width=0.33\linewidth,trim=0.5cm 1.5cm 0.5cm 2cm,clip=true]{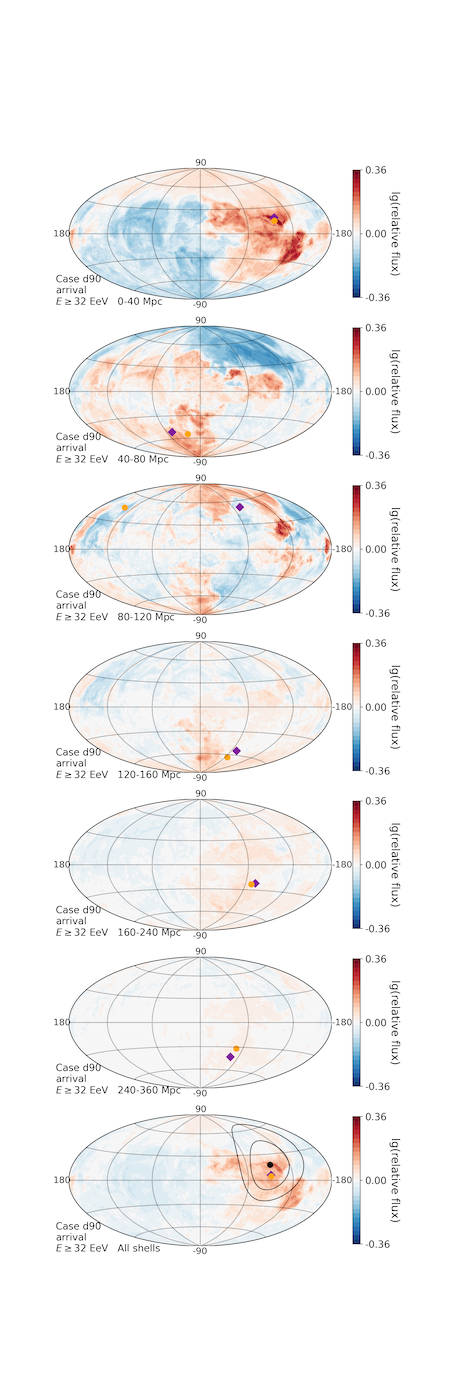}
\caption{Illumination maps (UHECR flux illuminating the Galaxy) and arrival maps (after propagation in the JF12 GMF model) for the model Case ``d90'' and energy bins $\geq 8$ EeV and $\geq 32$ EeV for different shells of distances.}
\label{fig:allshells_d90}
\end{figure*}

\begin{figure*}
\centering
    \centering
         \centering
         \includegraphics[width=0.33\linewidth,trim=0.5cm 1.5cm 0.5cm 2cm,clip=true]{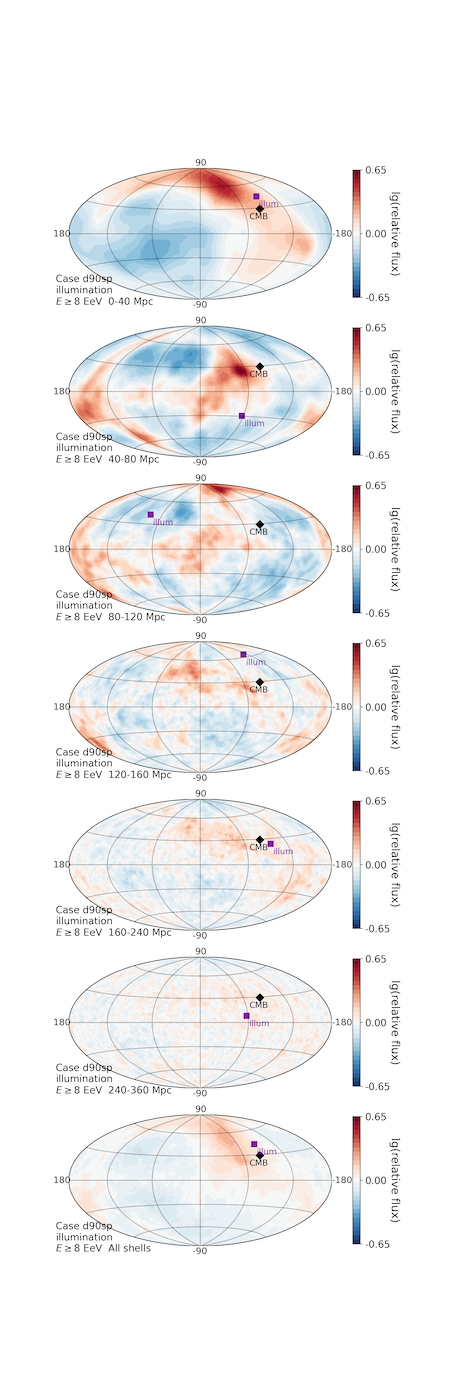}
         \includegraphics[width=0.33\linewidth,trim=0.5cm 1.5cm 0.5cm 2cm,clip=true]{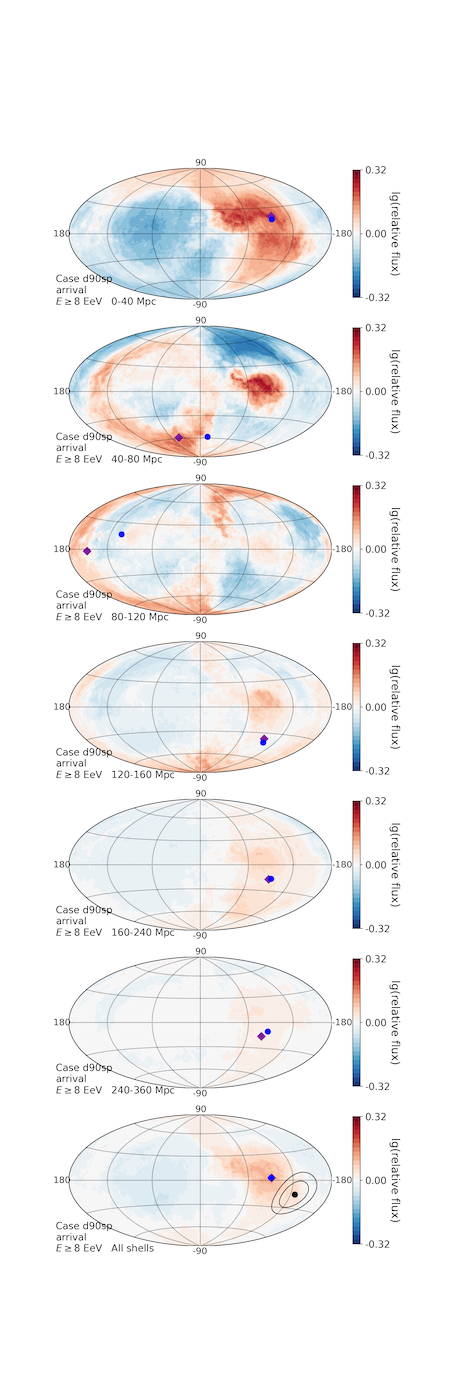}         
         \includegraphics[width=0.33\linewidth,trim=0.5cm 1.5cm 0.5cm 2cm,clip=true]{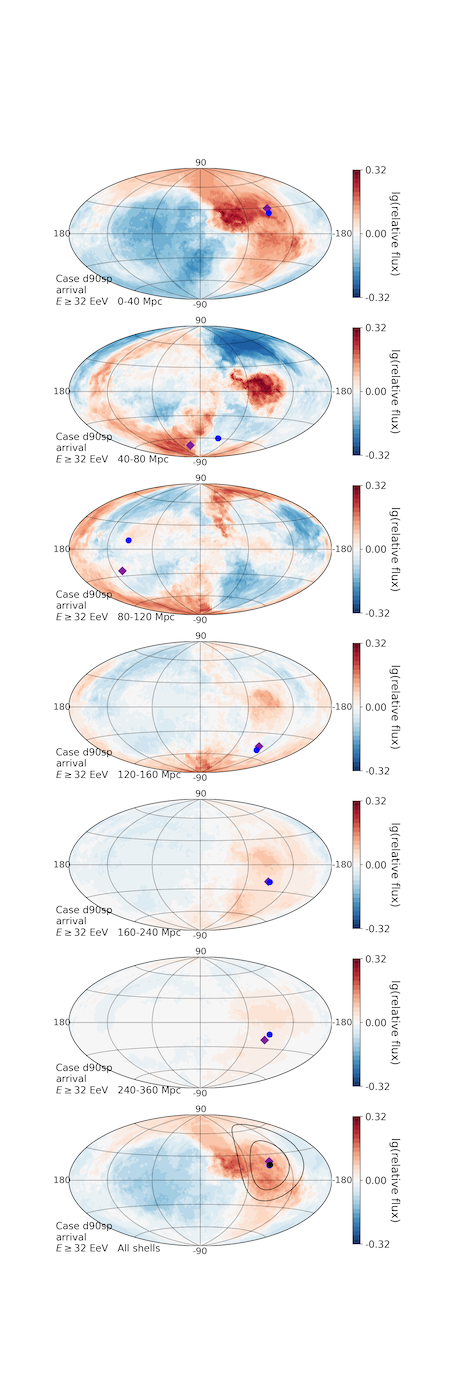}
\caption{Illumination maps (UHECR flux illuminating the Galaxy) and arrival maps (after propagation in the JF12 GMF model) for the model Case ``d90sp'' and energy bins $\geq 8$ EeV and $\geq 32$ EeV for different shells of distances.}
\label{fig:allshells_d90sp}
\end{figure*}

\begin{figure*}
\centering
    \centering
         \centering
         \includegraphics[width=0.33\linewidth,trim=0.5cm 1.5cm 0.5cm 2cm,clip=true]{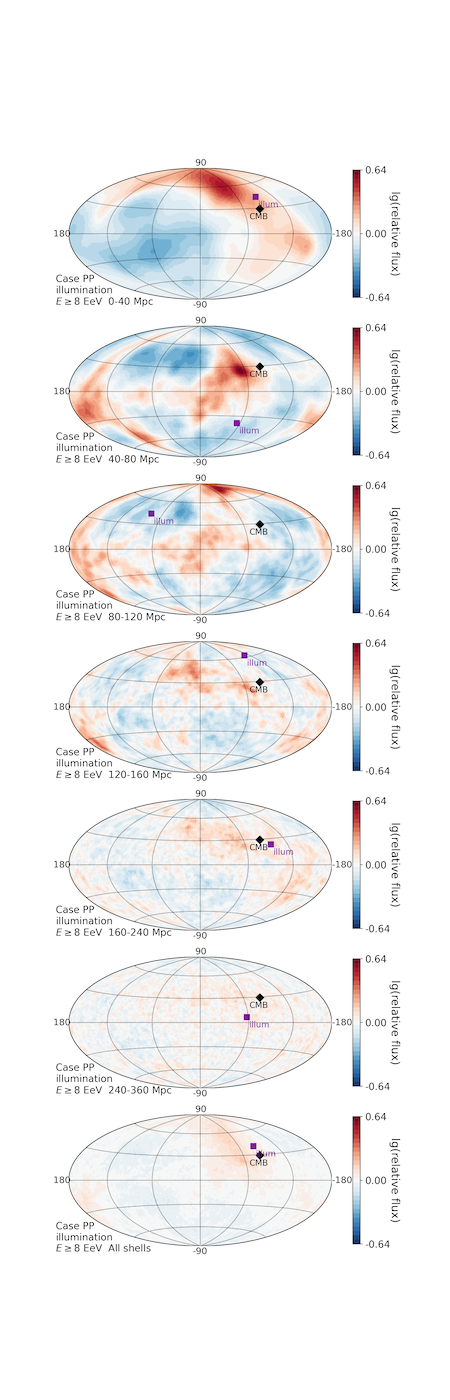}
         \includegraphics[width=0.33\linewidth,trim=0.5cm 1.5cm 0.5cm 2cm,clip=true]{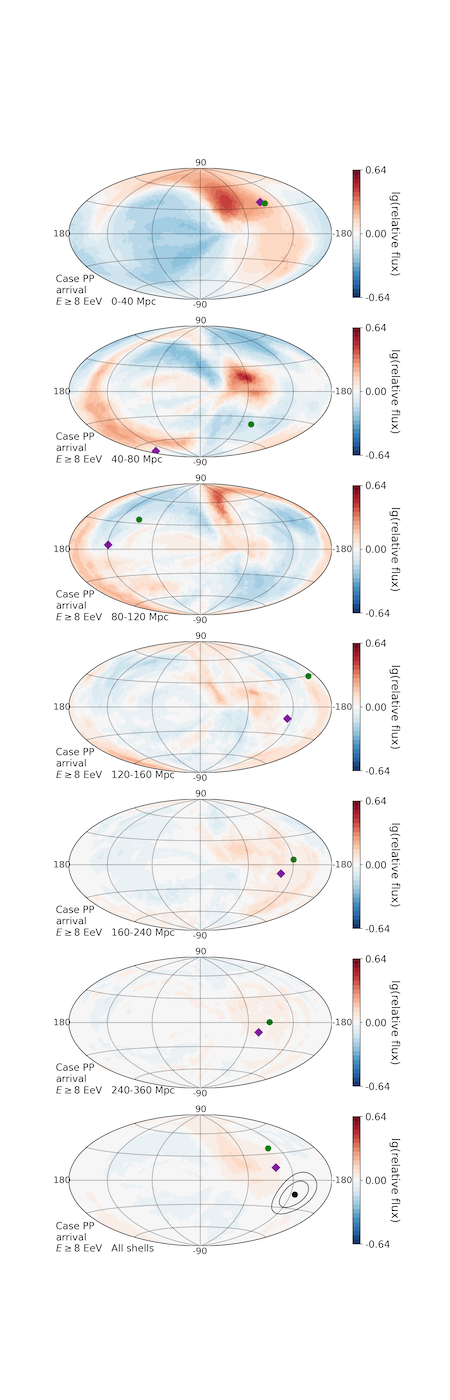}         
         \includegraphics[width=0.33\linewidth,trim=0.5cm 1.5cm 0.5cm 2cm,clip=true]{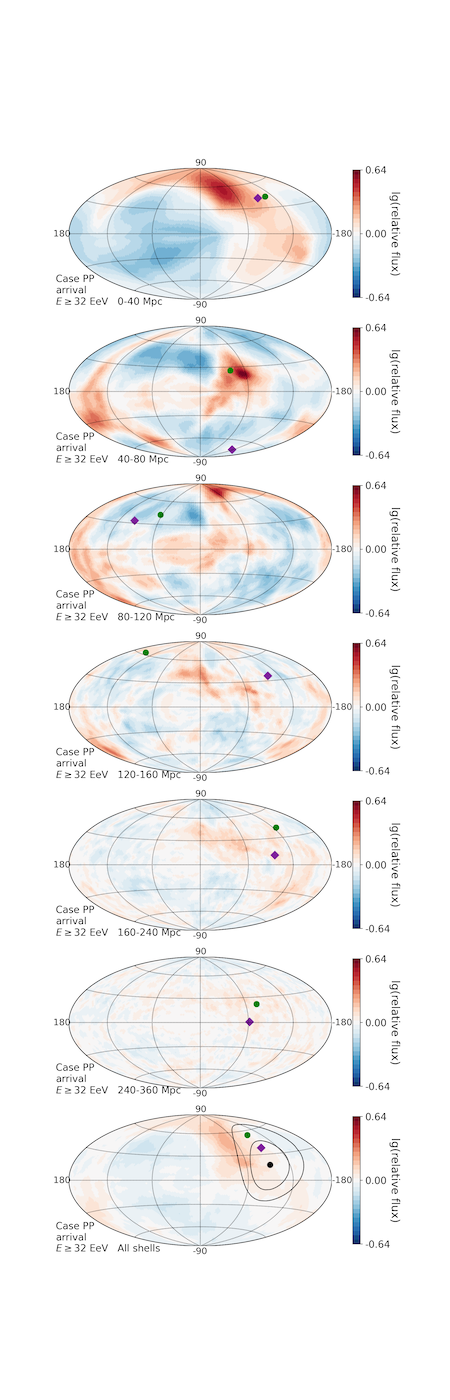}
\caption{Illumination maps (UHECR flux illuminating the Galaxy) and arrival maps (after propagation in the JF12 GMF model) for the model Case ``PP'' and energy bins $\geq 8$ EeV and $\geq 32$ EeV for different shells of distances.}
\label{fig:allshells_PP}
\end{figure*}

The left columns of Figs.~\ref{fig:allshells_SHasterisk}-\ref{fig:allshells_PP} show the illumination maps (UHECR flux illuminating the Galaxy) with unit weight from each shell, in the SH*, d90, d90sp and PP model cases. The center and right columns show the arrival maps (after propagation in the JF12 GMF model) in energy bins $\geq 8$ EeV and $\geq 32$ EeV for different shells of distances. The maps in the last row include cosmic rays from all distances with attenuation treatment.

The d90sp model uses the composition of the best-fitting SH* model, with the exponential attenuation treatment.  Comparing the first columns of Fig.~.~\ref{fig:allshells_d90} and Fig.~\ref{fig:allshells_d90sp} gives insight to the composition sensitivity of the illumination maps, since the d90 composition is heavier than that of SH*.  

\subsection{More Cases}
Tab.~\ref{tab:more_cases} summarize the results of additional cases to those listed in Tab.~\ref{tab:likelihood}. For the Cases SH*, SH$_{\rm E}$* and SH$_{\rm Q}$*, the best-fit parameters of the model are found by maximizing the likelihood $L$ (Eq.~\eqref{eq:likelihood}) for each assumption of HIM. The result leads to a composition that matches Sibyll2.3c the best for all three cases. To investigate what would be the likelihood to fit the anisotropy data with a UHECR composition inferred from $X_{\rm max}$ measurement with QGSJETII-04, we considered the Case SH$_{\rm forceQ}$*, where we maximize $\ln L(\langle\ln A\rangle\mid\boldsymbol{\Omega};\mathrm{HIM}) + \ln L(\sigma^2(\ln A)\mid\boldsymbol{\Omega};\mathrm{HIM})$ with composition parameter $\boldsymbol{\Omega}$ first, and then maximize $\ln L(\mathrm{dipole}\mid\boldsymbol{\Theta};\mathrm{source}) + \ln L(\mathrm{events}\mid\boldsymbol{\Theta};\mathrm{source})$ varying the other two parameters $D_{\rm EG,5EV}$ and $\lambda_{\rm G}$. Evidently, the fit to the anisotropy with a light composition is quite poor. The strength of the predicted LSS hotspot for our best-fitting description, case SH*, is weaker than the one observed. This could be the result of fitting to the arrival directions of all events above 38 EeV, which contains more information than just the strength of the hotspot, including the shape of the hotspot and the South Pole excess. 
To see whether a variation of the model parameters can give a similarly significant hotspot as observed, we carried out the exercise reported as Case SH(better hotspot)*, for which the objective function also includes the number of events inside the hotspot. The conclusion is that indeed the hotspot can be readily described, if it is included in the fitting, without significant damage to the fit to dipole components and events above 38 EeV. 

\begin{table*} 
\centering
\begin{footnotesize}
\begin{tabular}{|c|c|c|c|c|c|c|c|}
\hline
Cases & & Iso & SH* & SH$_{\rm E}$* & SH$_{\rm Q}$* & SH$_{\rm forceQ}$* & SH(better hotspot)* \\ \hline
\multirow{3}{*}{\rotatebox{90}{Model}} & Source model & Isotropic & LSS & LSS & LSS & LSS & LSS \\ 
& Distance weighting & --- & Sharp & Sharp & Sharp & Sharp & Sharp \\
& HIM & Sibyll2.3c & Sibyll2.3c & EPOS-LHC & QGSJET & QGSJET & Sibyll2.3c \\ \hline
\multirow{5}{*}{\rotatebox{90}{Likelihood}} & $\ln L(\mathrm{dipole}\mid\boldsymbol{\Theta};\mathrm{source})$ & -3.4 & 14.5 & 13.8 & 13.9 & -17.9 & 13.9 \\
& $\ln L(\mathrm{events}\mid\boldsymbol{\Theta};\mathrm{source})$ & 0 (Ref) & 11.1 & 10.9 & 11.0 & -11.3 & 9.6  \\
& $\ln L(\langle\ln A\rangle\mid\boldsymbol{\Omega};\mathrm{HIM})$ & 4.4 &  4.0 & 3.8 & -1.4 & 3.9 & 3.7 \\
& $\ln L(\sigma^2(\ln A)\mid\boldsymbol{\Omega};\mathrm{HIM})$ & -2.8 & -3.3  & -3.7 & -8.0 & -8.9 & -3.3  \\ \cline{2-8}
& Sum of $\ln L$ (Eq.~\eqref{eq:likelihood}) & -1.9 & 26.3 & 24.7 & 15.4 & -34.2 & 23.8  \\ \hline
\multirow{7}{*}{\rotatebox{90}{Best-fit parameters}} 
& $\lg D_{\rm EG,5EV}$ & --- & $2.79^{+0.60}_{-0.20}$  & $2.79^{+0.60}_{-0.20}$ & $2.79^{+0.60}_{-0.20}$ & 2.19 & 2.79 \\
& $\lg\lambda_{\rm G}$ & --- & $1.58^{+0.10}_{-0.08}$ & $1.58^{+0.13}_{-0.08}$ & $1.61^{+0.08}_{-0.10}$  & 2.00 & 1.53 \\
& $\langle\ln A\rangle_{\rm8-10\,EeV} $  & 2.02 & $1.76^{+0.19}_{-0.15}$ & $1.64^{+0.18}_{-0.16}$ & $1.62^{+0.15}_{-0.14}$ & 0.72 &  1.65\\
& $\langle\ln A\rangle_{\geq 40 {\rm\,EeV}} $ & 3.19 & $2.87^{+0.17}_{-0.10}$ & $2.84^{+0.13}_{-0.08}$ & $2.77^{+0.08}_{-0.07}$ & 1.89 & 2.68 \\
& $\sigma^2(\ln A)_{\rm8-10\,EeV} $ & 0.31 & $0.48^{+0.27}_{-0.20}$ & $0.47^{+0.27}_{-0.24}$ & $0.38^{+0.28}_{-0.18}$ & 0.49 & 0.57  \\
& $\sigma^2(\ln A)_{\geq 40 {\rm\,EeV}}$ & 0.19 & $0.09^{+0.32}_{-0.05}$ & $0.09^{+0.34}_{-0.05}$ &$0.05^{+0.13}_{-0.01}$ & 0.30 & 0.04 \\
\cline{2-8}
& $B_{\rm EG}$ if $\lambda_{\rm EG}=0.2$ Mpc  & --- & $0.32^{+0.08}_{-0.16}$ & $0.32^{+0.08}_{-0.16}$ & $0.32^{+0.08}_{-0.16}$ & 0.63 & 0.32 \\
\hline
\multirow{2}{*}{\rotatebox{90}{hotspot}} 
&\hspace*{-0.2in}\makecell{Number of events in $27^\circ$ circle\\centered at ($309.7^\circ,17.4^\circ$).  Obs=188} & $125^{+11}_{-11}$  & $154^{+12}_{-11}$ & $156^{+12}_{-12}$ & $160^{+12}_{-11}$ & $155^{+12}_{-12}$ & $168^{+12}_{-12}$ \\
\cline{2-8}
&\hspace*{-0.2in}\makecell{Li-Ma significance in $27^\circ$ circle\\centered at ($309.7^\circ,17.4^\circ$). Obs=5.6}& $0.0^{+1.0}_{-1.0}$ & $2.7^{+1.1}_{-1.0}$ & $2.8^{+1.1}_{-1.1}$ & $3.2^{+1.0}_{-1.0}$ & $2.7^{+1.1}_{-1.1}$ & $3.9^{+1.0}_{-1.0}$ \\
\hline
\end{tabular}
\end{footnotesize}
\caption{\small Summary of model parameters and results for cases in addition to those in Tab.~\ref{tab:likelihood}.  For Cases with asterisk (e.g. SH*), Eq.~\eqref{eq:likelihood} is maximized, i.e., models are fit to composition, dipole and events above 38 EeV.  The best-fit parameters are reported as the median with $1\sigma$ confidence levels (i.e. 16\% and 84\% percentiles). In the last two rows, the hotspot results are calculated from millions of mock data sets generated from the model arrival map above 40 EeV with the best-fit parameters. The confidence level represents the statistical uncertainty in the mock data datasets, and does not represent the uncertainty due to the uncertainty in best-fit parameters.} 
\label{tab:more_cases}
\end{table*}

\begin{table}
\centering
\begin{footnotesize}
\begin{tabular}{|c|c|c|c||c|c|c|c|}
\hline
{Observable} & {Auger} & {Auger} & {Auger and TA}  & {Case SH*} & {Case SH*} & {Case d90} & {Case d90}\\
& $l_{\rm max}=1$ & $l_{\rm max}=2$ & all multipoles & $l_{\rm max}=1$ & all multipoles & $l_{\rm max}=1$ & all multipoles \\
\hline
$d_x$ & $-0.008\pm0.009$ & $-0.004\pm0.012$ & $-0.007\pm0.011$ & -0.032 & -0.024 & -0.024 & -0.017 \\
$d_y$ & $0.059\pm0.009$ & $0.054\pm0.012$ & $0.042\pm0.011$ &  0.039 & 0.031 & 0.026 & 0.022\\
$d_z$ & $-0.028\pm0.014$ & $-0.011\pm0.035$ & $-0.026\pm0.019$ & -0.054 & -0.048 & -0.053 & -0.042\\
$Q$ &  & $0.032\pm0.014$ & & & 0.025 & & 0.020\\
\hline
\end{tabular}
\end{footnotesize}
\caption{\small Comparison of dipole and quadrupolar components $\geq 8$ EeV between observation and model. Since the Auger observatory has partial exposure coverage, the reconstruction these components needs the assumption that the anisotropy is purely dipolar ($l_{\rm max}=1$) or dipolar and quadrupolar ($l_{\rm max}=2$). The joint result of Auger+TA in the last column has full sky exposure coverage \citep{ICRC2019_AugerTA_anisotropy}.}
\label{tab:result_above8}
\end{table}

\subsection{Dipole and quadrupole components}\label{Appendix:dipole}
Table~\ref{tab:result_above8} compares the dipole and quadrupole components of the LSS model arrival map $\geq 8$ EeV with Auger observation \citep{Auger_4bin_2018,Auger_all_bins_2020,ICRC2019_AugerTA_anisotropy}

\subsection{Mock datasets above 38 EeV} \label{Appendix:lima}

\begin{figure}
\centering
\includegraphics[width=0.5\linewidth]{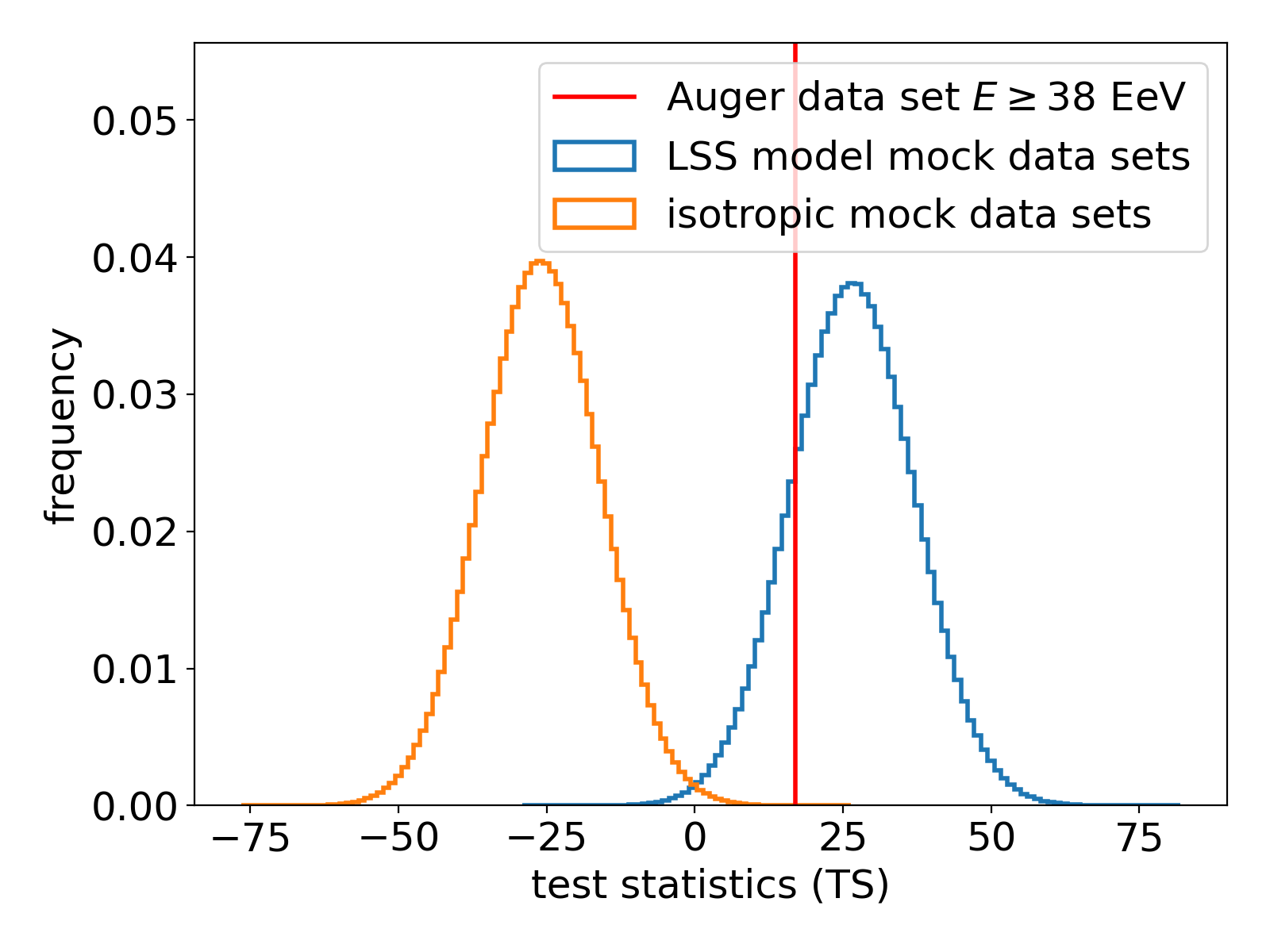}
\caption{Test statistics of model mock data sets, isotropy mock datasets and Auger data set.}
\label{fig:TS}
\end{figure}

By simulating mock Auger data sets for the LSS model with sharp cutoff treatment (Case SH*) and for an isotropic sky (Case Iso), we obtain the distribution of test statistics shown in Fig.~\ref{fig:TS}. The test statistic $\mathrm{TS} = 2\ln L(\mathrm{events}\mid\boldsymbol{\Theta};\mathrm{source}) = 22.1$, while the isotropic model is disfavored against the LSS sharp-cutoff model by 4.8 $\sigma$. This is to be compared \citet{Auger_starburst_gAGN} which found the events above 39 EeV are correlated with a starburst galaxy catalog, with $\mathrm{TS}=24.9$ and $4.0\sigma$ significance.

\subsection{Corner plots of parameters}

\begin{figure*}
\centering
\includegraphics[width=1\linewidth]{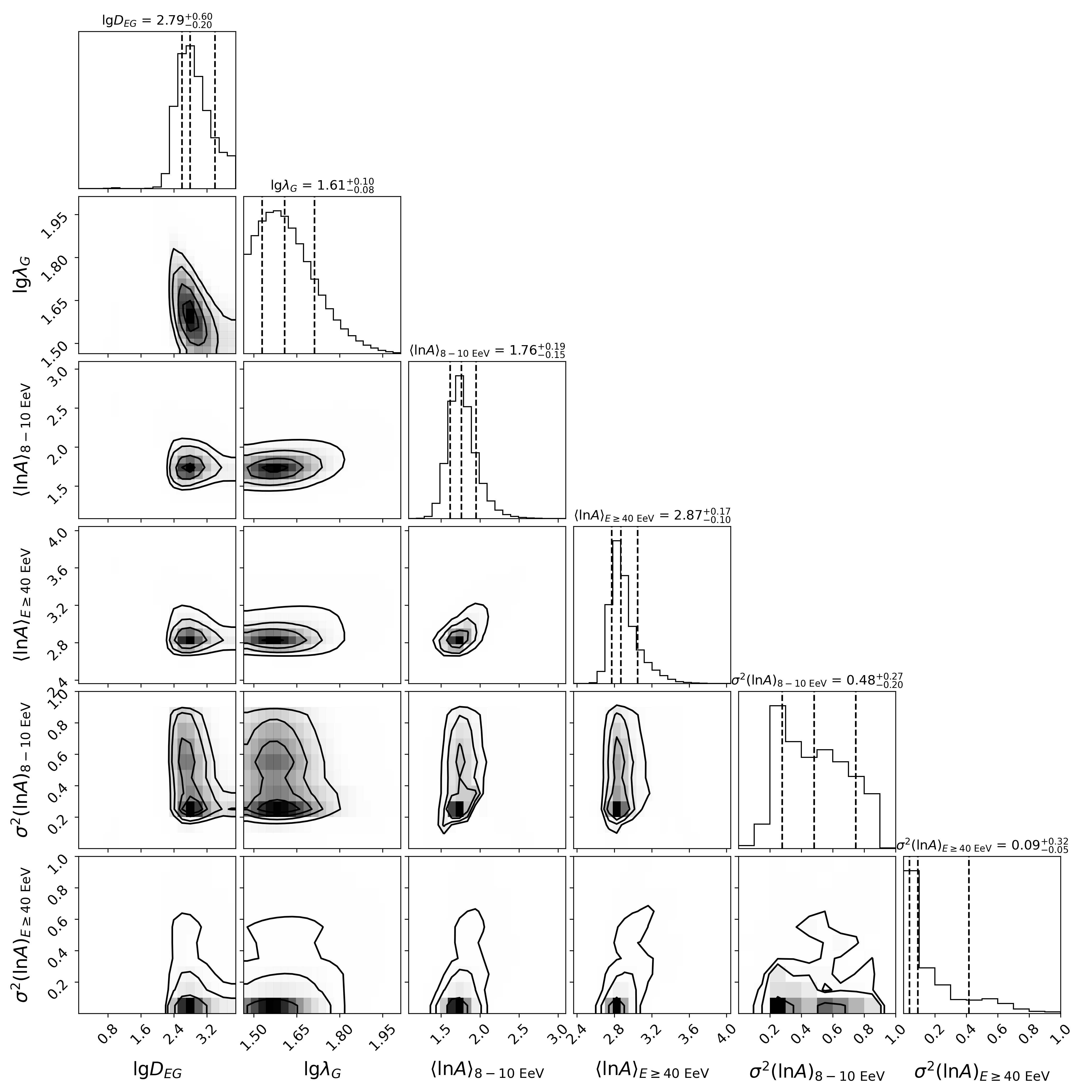}
\caption{The corner plot of the probability distribution of parameters, assuming Sibyll2.3c as the HIM and using the sharp-cutoff treatment (case SH*).}
\label{fig:corner_plot_Sibyll}
\end{figure*}

Figure~\ref{fig:corner_plot_Sibyll} completes Table~\ref{tab:likelihood} and shows the corner plot of the probability distribution of parameters, assuming Sibyll2.3c as the HIM. The 1D histograms on the diagonal show the probability distribution of each individual parameter, marginalized over all other five parameters. The 2D histograms off-diagnoal show the probability distribution of two parameters marginalized over all other four parameters. 

\subsection{Result with pure proton composition} \label{Appendix:pure_proton}
\begin{figure}
\centering
\includegraphics[width=0.45\linewidth]{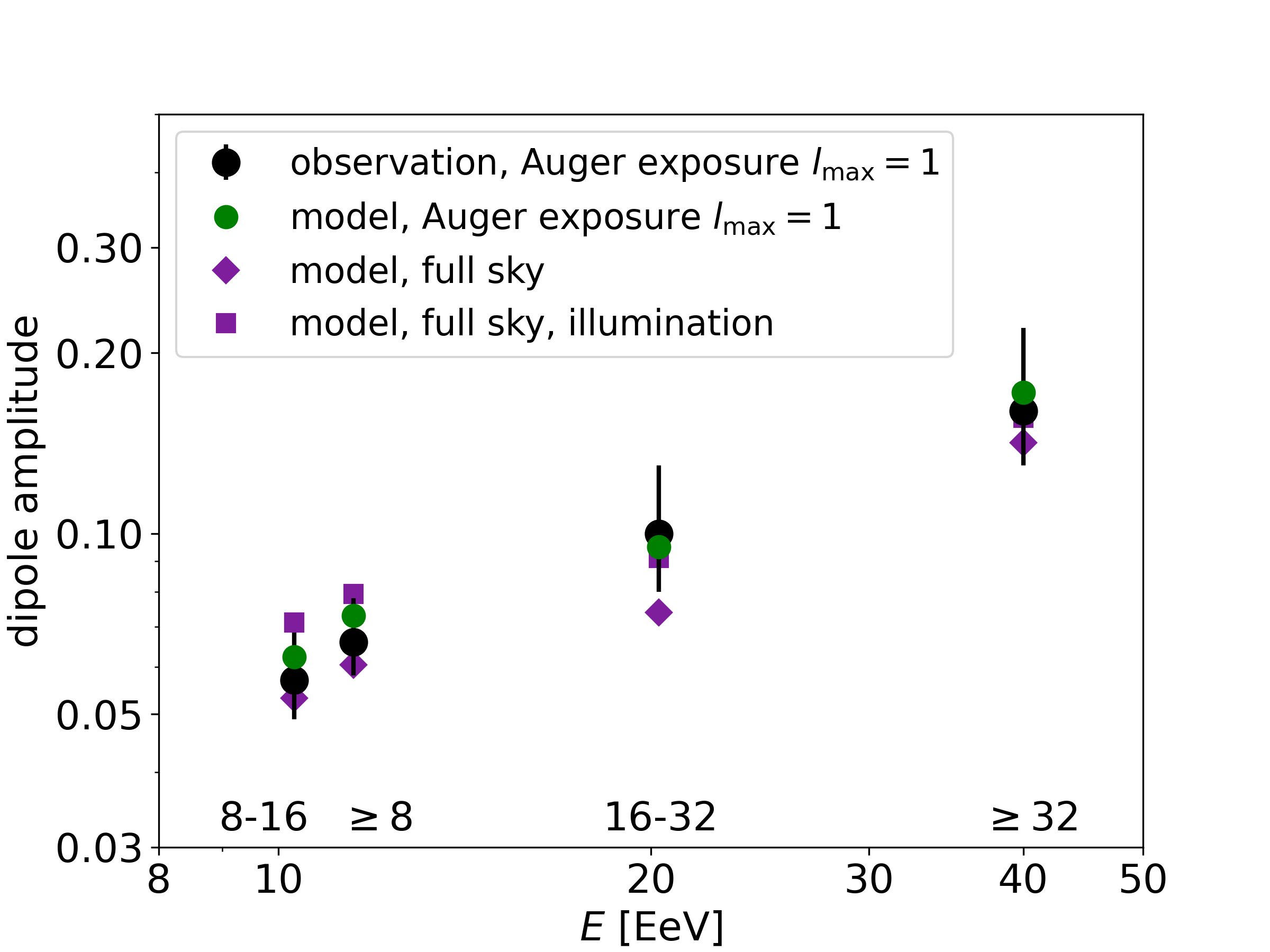}
\includegraphics[width=0.46\linewidth,trim=0cm 0cm 11cm 0cm,clip=true]{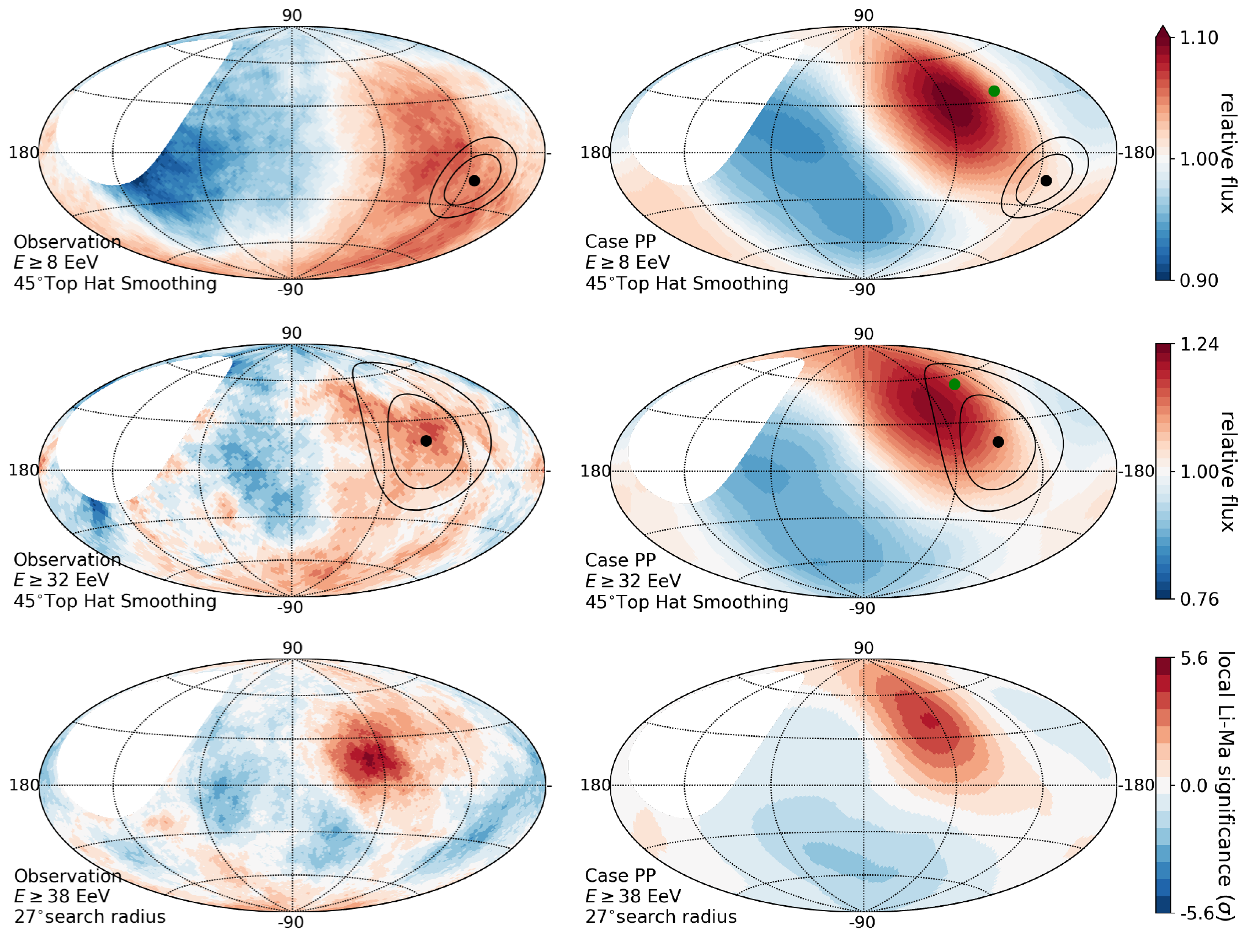}
\caption{The sky maps for the pure proton composition with exponential cutoff (Case PP) in the same format as Fig.~\ref{fig:dip_AM}}
\label{fig:pureproton}
\end{figure}

As shown in Table~\ref{tab:likelihood} Case PP, the pure proton composition gives a much poorer fit to the anisotropy than any other cases. This is illustrated in Fig.~\ref{fig:pureproton}, showing the best pure proton fit to the dipole anisotropy $\geq 8$ EeV. The left panel shows that the dipole amplitude can be in reasonable agreement but the arrival map excess is in an entirely wrong direction: in the northern hemisphere, far from the observed dipole direction \citep{Auger_Science_2017}. The small electric charge of the proton does not allow the required large deflection by the GMF, which occurs for the actual mixed composition. \citet{Ahlers:2017wpb}, using a simplified treatment, also concluded that pure proton composition is disfavored.

\end{CJK*}
\end{document}